\documentclass[12pt]{article}
\usepackage[latin1]{inputenc} 
\usepackage[english]{babel}
\usepackage{indentfirst}
\usepackage{fancyhdr}
\usepackage{graphicx}
\usepackage{newlfont}
\usepackage{amssymb} 
\usepackage{amsmath}
\usepackage{latexsym}
\usepackage{amsthm}
\usepackage{eucal}
\usepackage{eufrak}
\usepackage{bbold}
 \usepackage{tocbibind}
 \usepackage{newlfont}
 \usepackage{enumerate}
 \usepackage{listings}
 \usepackage{tikz}
 \usepackage{bbm}
\usepackage{color}
\usepackage[toc,page]{appendix}
\usepackage{hyperref}

\usepackage[square,sort&compress,numbers]{natbib}

\usepackage{color}

\numberwithin{equation}{section}

 \def \r {{\rho}}
 \def \br {{\bar{\rho}}}
 
  \def \tr {{\tilde{\rho}}}

\definecolor{mygreen}{RGB}{28,172,0} 
\definecolor{mylilas}{RGB}{170,55,241}

\theoremstyle{plain}                    
\newtheorem{theorem}{\bf Theorem}[section]

\newtheorem{proposition}[theorem]{ Proposition}
\newtheorem{example}[theorem]{\bf Example}

\newtheorem{assumption}[theorem]{\bf Assumption}

\newcommand*\samethanks[1][\value{footnote}]{\footnotemark[#1]}

\makeatletter
\newcommand*{\rom}[1]{\expandafter\@slowromancap\romannumeral #1@}
\makeatother

\numberwithin{equation}{section}

\author{Francesca Biagini\thanks{Workgroup Financial and Insurance Mathematics, Department of Mathematics, Ludwig-Maximilians Universit{\"a}t, Theresienstraße 39, 80333 Munich, Germany. Emails: biagini@math.lmu.de, meyer-brandis@math.lmu.de, mazzon@math.lmu.de.} \thanks{Secondary affiliation: Department of Mathematics, University of Oslo, Box 1053, Blindern, 0316, Oslo, Norway.} \and Andrea Mazzon\samethanks[1] \and Thilo Meyer-Brandis\samethanks[1]} 
 
 \title {Financial asset bubbles in banking networks}

\begin{document} 

\lstset{language=Matlab,%
    breaklines=true,%
    morekeywords={matlab2tikz},
    keywordstyle=\color{blue},%
    morekeywords=[2]{1}, keywordstyle=[2]{\color{black}},
    identifierstyle=\color{black},%
    stringstyle=\color{mylilas},
    commentstyle=\color{mygreen},%
    showstringspaces=false,
    numbers=left,%
    numberstyle={\tiny \color{black}},
    numbersep=9pt, 
    emph=[1]{for,end,break},emphstyle=[1]\color{red}, 
}

\maketitle

 \begin{abstract}
 We consider a banking network represented by a system of stochastic differential equations coupled by their drift.  We assume a core-periphery structure, and that the banks in the core hold a bubbly asset. The banks in the periphery have not direct access to the bubble, but can take initially advantage from its increase by investing on the banks in the core. Investments are modeled by the weight of the links, which is a function of the robustness of the banks. In this way, a preferential attachment mechanism towards the core takes place during the growth of the bubble. We then investigate how the bubble distort the shape of the network, both for finite and infinitely large systems, assuming a non vanishing impact of the core on the periphery. Due to the influence of the bubble, the banks are no longer independent, and the law of large numbers cannot be directly applied at the limit. This results in a term in the drift of the diffusions which does not average out, and that increases systemic risk at the moment of the burst. We test this feature of the model by numerical simulations. 
\end{abstract}


\section{Introduction}

Contagion mechanisms within a banking system and corresponding measurement and management of systemic risk have become central topics in macroprudential regulation in particular since the last financial crisis. The urge for the development of new quantitative methods to deal with these topics has triggered various lines of active research on systemic risk.

One stream of research aims at extending the traditional regulatory framework of monetary risk measures, that quantify the risk of financial institutions based on a stand alone basis, to multivariate systemic risk measures that take as a primitive the whole financial system. For an overview about this topic, see \citet{biaginisyst, biaginisyst2018},  \citet{bisias2012}, \citet{chen2013}, \citet{drapeau2016dual}, \citet{Feinstein2017}, \citet{Hoffmann2016, Hoffmann2017}, \citet{kromer2016} and references therein. 

Another popular ansatz to analyse systemic risk is based on explicit network models for the financial system and the study of potential default cascades due to various contagion affects. In the seminal work of \citet{Eisenberg} and its many extensions (see e.g. \citet{hurd2016contagion} and references therein) cascade processes in static, deterministic network models are analized by computing endogenously determined clearing/equilibrium payment vectors. Within the framework of random graph theory, cascade processes are studied in large financial random networks by means of law-of-large number effects in \citet{AminiMinca, Amini2012, Amini2013}, \citet{Detering2017, Detering2018, detering2015bootstrap} and \citet{hurd2016contagion}, and in finite random networks by  \citet{Elliott}, \citet{GaiKapadia}. 

The approach we present in this paper is placed within the theory of mean-field equations first introduced in the influential papers of \citet{McKean1, McKean2}. In recent years, this framework has been applied to the study of systemic risk in large financial networks where, contrary to the static network models mentioned above, the dynamic evolution of a network of interacting financial institutions is studied by means of a system of interacting diffusions. In this setting the diffusions represent e.g.~the wealth, monetary reserves, or other more general indicators for the health of financial institutions, and are tied together through a term in the drift that implies the network structure. A first simple model in this direction is given in \citet{Fouque1}, where a system of SDEs is proposed with dynamics
\begin{equation}\label{fouque}
dX^{i}_t=\frac{\lambda}{n}\sum_{j=1}^n (X^{j}_t - X^{i}_t)dt+\sigma dW_t^i, \quad 0 \le t < \infty, 
\end{equation}  
 where $W=(W_t^1,\dots,W_t^n)_{t \ge 0}$ is a standard $n$-dimensional Brownian motion and $\lambda, \sigma>0$.  
Here, the $X^{i}$ stand for the log-monetary reserves of banks, and the drift terms $\lambda(X^{j}_t - X^{i}_t)$ represent the connections between banks in the network. In this case, the borrowing and lending rate $\lambda$ is supposed to be the same for every couple of banks. When the network size $n$ grows towards infinity, it is a well-know result (see \citet{sznitman1991topics}) that due to law-of-large-number effects the diffusions in \eqref{fouque} converge towards their mean-field limit
 \begin{equation}\notag
d\bar{Y}^i_t=\lambda \left(\mathbb{E}[\bar{Y}_t] - \bar{Y}^i_t\right)dt+\sigma dW_t^i, \quad 0 \le t < \infty.
\end{equation} 

Thus, for large networks propagation of chaos applies and the evolution of the $X^i$ asymptotically de-couples due to averaging effects, which allows to asymptotically describe the complex system by a representative particle evolution. The simple model in \eqref{fouque} to study systemic risk has been generalized in various ways in a number of articles, see e.g. \citet{Carmona1, Carmona2} where mean-field games are considered, \citet{Fouque2} where the probability distributions of multiple default times is approximated, \citet{Garnier, Garnier2} and \citet{BattistonS} where a tradeoff between individual and systemic risk in a banking network is described, and \citet{Kluppelberg, Kley} where partial mean-field limits are studied.  \\
In this paper the main objective is to extend the model in \eqref{fouque} such that the effect of a financial speculation bubble on the evolution of the network and the evolving systemic risk can be studied. It is a common understanding that bubbles are intimately connected with financial crises, and many historical crises indeed originated after the burst of a bubble (e.g. the Great Depression of the 1930s and the financial crisis of 2007-2008). This causality is investigated for example in \citet{Brunnermeier2008} and statistically confirmed in \citet{brunnermeier2015bubbles}. However, it seems that literature on mathematical models that deal with this question is very scarce. 

We here take a first step towards filling this gap and specify a model for the network of \textit{financial robustness} of the institutions, introduced by \citet{BattistonS} and \citet{HullWhite} as an indicator of agent's creditworthiness or distance to default and also considered in \citet{Kley}, by a system of coupled diffusions. The banks affect each other's robustness by being financially exposed to each other, for example because of cross-holdings, which results in a coupling of the drift terms. Following the setting in \citet{Battiston}, we then assume that a fixed number of banks are directly investing in a bubble that affects their financial robustness. The remaining banks have the possibility to participate in the bubble by investing in the bubble banks. This results in a typical core/periphery structure for financial networks, where here the core is formed by the banks holding the bubble. Contrarily to the literature on mean-field models mentioned above, where the coupling drift rates representing the weighted network connections are constant, we allow for heterogeneity of the drift rates in that they depend on the robustness of the institution. More precisely, the rates depend on the robustness of the attracting institution with a delay $\delta>0$,  where the delay reflects the fact that the banks' investment do not immediately react to changes in the system.  This results in a preferential attachment mechanism where the attractiveness of a node does not depend on its degree, but on its  ``fitness'', as  proposed by \citet{BianconiBarabasi}. Due to this behavior, the bubble causes a distortion in the network evolution: during the expanding phase of the bubble, the network structure shifts towards an increasingly intense and centralized connectivity due to the strong growth of the bubbly banks' robustness, which then causes instability in case the bubble bursts. 

We then study the behaviour of the system when its size gets large. More precisely, we let the number of periphery banks go towards infinity, but keep the number of core banks holding the bubble constant and assume that their impact on the system does not vanish when the total number of banks goes to infinity. In this way the bubble produces a common stochastic source in the system that does not not average out even for large networks. Our main result then determines a partial mean-field limit for the system where the influence of the bubble is represented via stochastic interaction with the core banks even in the limit. Because of this term, also the banks in the periphery are affected by a potential bubble burst. This effect is amplified by the impossibility to immediately desinvest when the robustness of some banks decreases due to the delay $\delta$. We also refer to \citet{Kluppelberg} where the authors investigate partial mean-field limits in a different setting, without taking into account the delay and the influence of the bubble.

%

%

The remaining part of the paper is organized as follows. In Section \ref{model} we introduce our model and some technical results. In Section \ref{seclimit} we define the limit system and prove a convergence result, whereas in Section \ref{simulations} we perform Monte Carlo simulations both in the finite and in the limit systems in order to numerically investigate the impact of the bubble on systemic risk.

\section{The model}\label{model}

Let  $(\Omega, \cal{F}, \mathbb{F},P)$ be a filtered probability space endowed with a $(m+n+2)$-dimensional Brownian motion $\bar{W}=(W^1_t,\dots,W^n_t,W^{B,1}_t,\dots,W^{B,m}_t,B_t^1,B_t^2)_{t \ge 0},$ $m, n \in \mathbb{N}$, where $\mathbb{F}=(\cal{F}_t)_{t \in \mathbb{R}^+}$ is the natural filtration of $\bar{W}$. 
We consider a network of $m+n$ banks, consisting of $m$ banks holding a bubbly asset in their portfolio (also referred to as \emph{core}), and $n$ banks that do not directly hold the bubbly asset (also referred to as \emph{periphery}). \\
By following a similar approach as in \citet{Kley}, we model the robustness of the banks in the system. This coefficient dynamically evolves and represents a measure of how healthy a bank remains in stress situations. Let  $\rho^{i,n}=(\rho^{i,n}_t)_{t \ge 0}$, $i=1,\dots n$, and $\rho^{k,B}=(\rho^{k,B}_t)_{t \ge 0}$, $k = 1, \dots, m$, be the robustness of banks not holding and holding the bubble, respectively. We assume that they satisfy the following system of stochastic differential delay equations (SDDEs) for $t \ge \delta$, $\delta >0$,
 \begin{align}
d\rho_t^{i,n}=&\left(\frac{1}{n-1}\sum_{j=1, j \ne i}^n f^P(\rho_{t-\delta}^{j,n}-A^n_{t-\delta}) (\rho_t^{j,n}-A_t^n) + \frac{1}{m} \sum_{k=1}^m f^B(\rho_{t-\delta}^{k,B}-A^n_{t-\delta})\ (\rho_t^{k,B}-A_t^n)\right)\notag \\&+\lambda(A_t^n-\rho_t^{i,n})dt +\sigma_1 dW_{t}^i, \label{rhonobubble} \\
d\rho_t^{k,B}=&\left(\frac{1}{n}\sum_{i=1}^n f^P(\rho_{t-\delta}^{i,n}-A^n_{t-\delta}) (\rho_t^{i,n}-A_t^n) + \frac{1}{m-1} \sum_{\ell=1,\ell \ne k}^m  f^B(\rho_{t-\delta}^{\ell,B}-A^n_{t-\delta}) (\rho_t^{\ell,n}-A_t^n)\right)dt\notag \\&+\lambda(A_t^n-\rho_t^{k,B})dt+\sigma_2dW_t^{k,B}+d\beta_t,  \label{rhobubble} 
\end{align}
 where $\lambda>0$, $\sigma_1>0$, $\sigma_2>0$ and 
 \begin{equation}\label{average}
 A_t^n=\frac{1}{m+n}\left(\sum_{r=1}^n \rho_t^{r,n}+\sum_{h=1}^m \rho_t^{h,B}\right), \quad t \ge \delta, 
 \end{equation}
 is the mean of the robustness of all the banks in the network at time $t$.
 For $t \in [0,\delta)$, we assume that  $(\rho_s^{i,n})_{s \in [0,\delta)}$, $(\rho_s^{k,B})_{s \in [0,\delta)}$, $i=1,\dots,n$, $k=1,\dots,m$, satisfy (\ref{rhonobubble})-(\ref{rhobubble}) with $\delta=0$, by following the approach of \citet{Mao}. We also suppose that  $\rho_0^{i,n}=\rho_0>0$ for all $i=1,\dots,n$. \\
%
%
The process $\beta=(\beta_t)_{t \ge 0}$ in \eqref{rhobubble} represents the influence of the asset price bubble on the robustness of core banks and has dynamics 
\begin{equation}\label{dbeta}
d\beta_t=\mu_tdt+\sigma_B dB_t^1, \quad t \ge 0,
\end{equation}
where $\sigma_B>0$ and $\mu$ is an adapted process satisfying 
\begin{equation}\label{dMu}
d\mu_t = \tilde{b}(\mu_t)dt+\tilde{\sigma}(\mu_t)dB^2_t, \quad t  \ge 0, 
\end{equation} 
where $\tilde{b}$, $\tilde{\sigma}$ fulfill the usual Lipschitz and sublinear growth conditions such that
there exists a unique solution
of (\ref{dMu}) , satisfying 
\begin{equation}\label{mu}
\int_{0}^t \mathbb{E}[|\mu_s|^2]ds<\infty, \quad  \quad 0 \le t < \infty.
\end{equation} 
Later on in Section~\ref{simulations} we will specify a concrete model for the bubbly evolution in \eqref{dbeta}.

The interdependencies of the banks' robustness and corresponding contagion effects are specified through the drifts in \eqref{rhonobubble} and \eqref{rhobubble}. The term $\lambda(A_t^n-\rho_t^{i,n})$ represents an attraction of the individual robustness towards the average robustness of the system with rate $\lambda$ as in the classical mean-field model \eqref{fouque}. In addition to the homogeneous average term, we introduce the terms of type $f^P(\rho_{t-\delta}^{j,n}-A^n_{t-\delta}) (\rho_t^{j,n}-A_t^n)$ and $f^B(\rho_{t-\delta}^{k,B}-A^n_{t-\delta})\ (\rho_t^{k,B}-A_t^n)$ that represent a robustness-dependent evolution of the network connectivity: for typically positive and increasing $f^B$ and $f^P$, bank $i$ is the more connected to bank $j$ the higher bank $j$'s robustness is above the average. In this way, the evolution of the bubble alters the connectivity structure of the network according to a model of preferential attachment. Moreover, the propensity of a node $i$ to attract future links not only depends on the current level of robustness of $i$, but also on the robustness of the banks already connected to $i$. This induces a form of \textit{preferential preferential attachment}, which creates a strong clustering effect. This change in network structure then comes along with an increasing systemic risk and instability in case the bubble burst, as noted by \citet{Battiston}. Further we introduce the delay $\delta > 0$ to reflect the fact that the bank i's investment decisions does not immediately react to changes in bank j's robustness. Note that when there are no bubble banks and $f^P=\lambda$, the system \eqref{rhonobubble}-\eqref{rhobubble} collapses to the basis mean-field model in \eqref{fouque}. \\
 We assume the following hypothesis on $f^B$ and $f^P$.
\begin{assumption}\label{f}
The functions $f^B,f^P: (\mathbb{R},\mathcal{B}(\mathbb{R})) \to (\mathbb{R}^+,\mathcal{B}(\mathbb{R}^+))$ are measurable, Lipschitz continuous and such that also the functions $F^B(x):=xf^B(x)$, $F^P(x):=xf^P(x)$, $x \in \mathbb{R}$, are Lipschitz continuous, i.e.
\begin{equation}\label{eq1f}
|f^{\ell}(x)-f^{\ell}(y)| \le K_1|x-y|, \quad x,y \in \mathbb{R}, \quad \ell=B,P,
\end{equation}
and
\begin{equation}\label{eqf}
|xf^{\ell}(x)-yf^{\ell}(y)| \le K
_2|x-y|, \quad x,y \in \mathbb{R}, \quad \ell=B,P,
\end{equation}
with $0<K_1,K_2<\infty$.
\end{assumption}
Note that (\ref{eqf}) implies that $f^B$ and $f^P$ are bounded, since if $f(x)x$ is Lipschitz  then
\begin{equation}\label{fK}
|f(x)x| = |f(x)x-f(0) \cdot 0| \le K_2|x|.
\end{equation}

\begin{example}\label{exf}
We have that $f(x)=1+2\arctan(x)/\pi$ satisfies Assumption \ref{f}: $f$ takes values in $[0,2]$, and both $f$ and $F(x)=xf(x)$ are Lipschitz, because they have bounded derivative.\\ In particular, $f$ is increasing, so that if $\rho_t^j>\rho_t^i$ then the link towards $j$ is bigger then the link towards $i$. If the robustness $\rho_t^j$ of bank $j$ is equal to the average $A_t^n$ in \eqref{average}, then the link towards bank $j$ has weight $f(0)=1$, if $\rho_t^j>A_t^n$  the link has weight bigger than $1$ and if $\rho^j_t<A^n_t$ the link has weight less than $1$. If all the banks have the same robustness, we have an homogenous network, where all the links have weight equal to $1$. \\ Furthermore, any constant function clearly satisfies Assumption \ref{f}. For such a choice, we have a static and homogenous network.
\end{example}



\begin{proposition}\label{promex}
Under Assumption \ref{f}, for every $\delta \ge 0$ there exists a unique strong solution for the system of SDEs (\ref{rhonobubble})-(\ref{rhobubble}). Moreover, it holds
\begin{align}
&\sup_{0 \le s \le t}\mathbb{E}[|\rho_s^i|^2]<\infty, \quad  0 < t <\infty,\quad i=1,\dots,n, \label{intrhoi}\\
&\sup_{0 \le s \le t}\mathbb{E}[|\rho^{k,B}_s|^2]<\infty, \quad  0 < t <\infty,\quad k=1,\dots,m. \label{intrhok}
\end{align}
\end{proposition}
\textit{Proof.} Suppose by simplicity $\lambda=1$. We start by proving existence and uniqueness of the strong solution of (\ref{rhonobubble})-(\ref{rhobubble}) when $\delta=0$. In this case we can write the system of SDEs given by (\ref{rhonobubble}),(\ref{rhobubble}) and (\ref{dMu}) as an $(m+n+1)$-dimensional SDE
\begin{equation}\label{eqX1}
dX_t=b(X_t)dt+\sigma(X_t)d\bar{W}_t, \quad t \ge 0,
\end{equation}
where
\begin{equation}\label{eqb1}
b(x)=\left( \begin{array}{c}\frac{1}{n-1}\sum_{j=2}^n f^P(x_j-\bar{x})(x_j-\bar{x})+\frac{1}{m}\sum_{k=n+1}^{m+n} f^B(x_k-\bar{x})(x_k-\bar{x})+\bar{x}-x_1,  \\
\vdots \\
\frac{1}{n-1}\sum_{j=1}^{n-1} f^P(x_j-\bar{x})(x_j-\bar{x})+\frac{1}{m}\sum_{k=n+1}^{m+n} f^B(x_k-\bar{x})(x_k-\bar{x})+\bar{x}-x_n \\ 
\frac{1}{n}\sum_{j=1}^{n} f^P(x_j-\bar{x})(x_j-\bar{x})+\frac{1}{m-1}\sum_{k=n+2}^{m+n} f^B(x_k-\bar{x})(x_k-\bar{x})+\bar{x}-x_{n+1}  \\
  \vdots \\
\frac{1}{n}\sum_{j=1}^{n} f^P(x_j-\bar{x})(x_j-\bar{x})+\frac{1}{m-1}\sum_{k=n+1}^{m+n-1} f^B(x_k-\bar{x})(x_k-\bar{x})+\bar{x}-x_{m+n} \\ 
\tilde{b}(x_{m+2}) \\ \end{array}\right),
\end{equation}
with $\bar{x}=\frac{1}{m+n}\sum_{i=1}^{m+n}x_i$.  Here $\sigma(x)$ is a $(n+m+1) \times (n+m+1)$ block matrix of the form
\begin{equation}\label{matsigma11}
\sigma(x)= \left( \begin{array}{ccc}
\Sigma_1(x) & 0 & 0 \\
0 & \Sigma_2(x) & 0 \\
0 & 0 &  \tilde{\sigma}(x_{m+2})\\
 \end{array} \right),
\end{equation}
where $\Sigma_1(x)$ is a $n \times n$ diagonal matrix with diagonal $(\sigma_1,\dots,\sigma_1)$ and $\Sigma_2(x)$ is the $m \times (m+1)$ matrix 
\begin{equation}\notag
\Sigma_2(x)= \left( \begin{array}{cccccc}
\sigma_2 & 0 & \dots & 0 & \sigma_B  \\
0 & \sigma_2 & \dots & 0 & \sigma_B  \\
\vdots & \vdots & \ddots & 0 & \sigma_B  \\
0 & 0 & \dots & \sigma_2 & \sigma_B \\
 \end{array} \right).
\end{equation}
We use Theorem 9.11 in \citet{Pascucci2011} to prove existence and uniqueness of the strong solution of (\ref{eqX1}), and that the second moments of the solution are finite. To this purpose, we show that $b(\cdot)$ and $\sigma(\cdot)$ defined in \eqref{eqb1} and \eqref{matsigma11}, respectively, are Lipschitz continuous in $x$ and that there exists some $C$ such that 
$$
\lVert \sigma(x) \rVert^2 + \lVert b(x) \rVert^2 \le C (1+ \lVert x \rVert^2).
$$
We begin by proving the first condition. The Lipschitz property clearly holds for $\sigma(\cdot)$, since $\tilde{\sigma}(\cdot)$ is Lipschitz  by hypothesis.
 Given $x=(x_1,\dots,x_{m+n}), x'=(x'_1,\dots,x'_{m+n}) \in \mathbb{R}^{m+n}$, we show that there exists a constant $\bar{K} \in (0,\infty)$ such that 
 $$\lVert b(x)-b(x')\rVert \le \bar{K} \lVert x-x'\rVert.$$
 For the first entry of (\ref{eqb1}) we have
\begin{align}
|b_1(x)-b_1(x')|  \le &  \frac{1}{n-1} \sum_{j=2}^{n} |f^P(x_j-\bar{x})(x_j-\bar{x})-f^P(x_j'-\bar{x}')(x_j'-\bar{x}')| \notag \\ &+ \frac{1}{m} \sum_{k=n+1}^{m+n} |f^B(x_k-\bar{x})(x_k-\bar{x})-f^B(x_k'-\bar{x}')(x_k'-\bar{x}')|+|\bar{x}-\bar{x}'|+|x_1-x_1'| \notag,
\end{align}
and by Assumption \ref{f} we have
\begin{align}
|b_1(x)-b_1(x')|&\le   K_2\frac{1}{n-1} \sum_{j=2}^{n} |(x_j-\bar{x})-(x_j'-\bar{x}')| +K_2\frac{1}{m} \sum_{k=n+1}^{m+n} |(x_k-\bar{x})-(x_k'-\bar{x}')|\notag \\ &\quad +|\bar{x}-\bar{x}'|+|x_1-x_1'|
\notag \\
& \le   K_2\left(\frac{1}{n-1} \sum_{j=2}^{n} |x_j-x_j'|+ \frac{1}{m}\sum_{k=n+1}^{m+n} |x_k-x_k'|\right)+(2K_2+1)|\bar{x}-\bar{x}'|+|x_1-x_1'|\notag \\
& \le  K_2\left(\frac{1}{n-1} \sum_{j=2}^{n} |x_j-x_j'|+\frac{1}{m} \sum_{k=n+1}^{m+n} |x_k-x_k'|\right)+\frac{2K_2+1}{m+n}\sum_{i=2}^{m+n}|x_i-x'_i|\notag \\ &\quad +|x_1-x_1'|.\notag
\end{align}
Then, since for $z_1,\dots,z_N \in \mathbb{R}$ it holds $\left(\sum_{i=1}^N |z_i|\right)^2 \le N\sum_{i=1}^N |z_i|^2$, we have
\begin{align}
 |b_1(x)-b_1(x')|^2  & \le  6(m+n) \left( (K_2)^2\left(\frac{1}{(n-1)^2} \sum_{j=2}^{n} |x_j-x_j'|^2+ \frac{1}{m^2}\sum_{k=n+1}^{m+n} |x_k-x_k'|^2\right)\right) \notag \\& \quad+6(m+n)\left((2K_2+1)^2\frac{1}{(m+n)^2}\sum_{i=1}^{m+n}|x_i-x'_i|^2+|x_1-x_1'|^2\right)\notag \\
& \le C_1 \lVert x-x' \rVert ^2, \notag 
\end{align}
for a suitable constant $C_1>0$.
 Similarly,
\begin{equation}\notag
|b_{i}(x)-b_{i}(x')|\le C_{i} \lVert x-x' \rVert ^2, \quad 2 \le i \le m+n,
\end{equation}
for a suitable constant $C_i>0$, whereas
\begin{equation}\notag
|b_{m+2}(x)-b_{m+2}(x')|= |\tilde{b}(x_{m+2})-\tilde{b}(x'_{m+2})| \le K_{\mu} |x_{m+2}-x'_{m+2}|,
\end{equation}
where $K_{\mu}$ is the Lipschitz constant for the function $\tilde{b}(\cdot)$ in (\ref{dMu}).
Then we obtain
\begin{align}\label{lipschitz1}
\lVert b(x)-b(x') \rVert^2 = \sum_{i=1}^{m+n+1} |b_{i}(x)-b_{i}(x')|^2 \le \left(\sum_{i=1}^{m+n+1}C_i + K_{\mu}^2\right) \lVert x-x' \rVert^2.  
\end{align}
The second condition, i.e. 
\begin{equation}\label{linear1}
\lVert \sigma(x) \rVert^2 + \lVert b(x) \rVert^2 \le C (1+ \lVert x \rVert^2), 
\end{equation}
for some $C>0$, holds because of Assumption \ref{f} and the hypothesis on $\tilde{\sigma}(\cdot)$.\\
Inequalities (\ref{intrhoi}) and (\ref{intrhok}) then follow by Theorem 9.11 in \citet{Pascucci2011}, and in particular by estimation (\ref{pascucci}) in the Appendix.   \\
When $\delta>0$, equation (\ref{eqX1}) becomes
\begin{equation}\label{eqXdelayed1}
dX_t=\bar{b}(X_t,X_{t-\delta})dt+\bar{\sigma}(X_t,X_{t-\delta})dW_t, \quad t \ge \delta,
\end{equation}
where $\bar{\sigma}(x,y)=\sigma(x)$ as in (\ref{matsigma11}) and 
\begin{equation}\notag
b(x,y)=\left( \begin{array}{c}\frac{1}{n-1}\sum_{j=2}^n f^P(y_j-\bar{y})(x_j-\bar{x})+\frac{1}{m}\sum_{k=n+1}^{m+n} f^B(y_k-\bar{y})(x_k-\bar{x})+\bar{x}-x_1,  \\
\vdots \\
\frac{1}{n-1}\sum_{j=1}^{n-1} f^P(y_j-\bar{y})(x_j-\bar{x})+\frac{1}{m}\sum_{k=n+1}^{m+n} f^B(y_k-\bar{y})(x_k-\bar{x})+\bar{x}-x_n \\ 
\frac{1}{n}\sum_{j=1}^{n} f^P(y_j-\bar{y})(x_j-\bar{x})+\frac{1}{m-1}\sum_{k=n+2}^{m+n} f^B(y_k-\bar{y})(x_k-\bar{x})+\bar{x}-x_{n+1}  \\
  \vdots \\y
\frac{1}{n}\sum_{j=1}^{n} f^P(y_j-\bar{y})(x_j-\bar{x})+\frac{1}{m-1}\sum_{k=n+1}^{m+n-1} f^B(y_k-\bar{y})(x_k-\bar{x})+\bar{x}-x_{m+n} \\ 
\tilde{b}(x_{m+2}) \\ \end{array}\right).
\end{equation}By Theorem 3.1 in \citet[chapter 5]{Mao}, to prove existence and uniqueness of the solution it suffices to show that the linear growth condition 
\begin{equation}\label{lineardelayed1}
\lVert \bar{b}(x,y) \rVert ^2 \le C(1+ \lVert x \rVert ^2 + \lVert y \rVert ^2)
\end{equation}
holds and that $\bar{b}$ is Lipschitz in the variable $x$ uniformly in $y$, i.e. that there exists a constant $\tilde{K} \in (0,\infty)$ such that  
\begin{equation}\label{lipschitzdelayed1}
\lVert \bar{b}(x,y)-\bar{b}(x',y) \rVert ^2 \le \tilde{K} \lVert x- x' \rVert ^2
\end{equation}
for all $y \in \mathbb{R}$, $x, x' \in \mathbb{R}^{m+n}$. Property (\ref{lineardelayed1}) can be proven by computations similar to the ones used for showing (\ref{linear1}). For the Lipschitz condition we have
 \begin{align}
|\bar{b}_1(x,y)-\bar{b}_1(x',y)| \le & \frac{1}{n-1} \sum_{j=2}^{n} |f^P(y_j-\bar{y})||(x_j-\bar{x})-(x_j'-\bar{x}')| \notag \\ &+ \frac{1}{m} \sum_{k=n+1}^{m+n} |f^B(y_k-\bar{y})||(x_k-\bar{x})-(x_k'-\bar{x}')|+|\bar{x}-\bar{x}'|+|x_1-x_1'|.\notag
 \end{align}
Hence, as $f^B$ and $f^P$ are bounded by $K_2$, the computations to show (\ref{lipschitzdelayed1}) are identical to the ones for (\ref{lipschitz1}). \\
 In order to prove (\ref{intrhoi}) and (\ref{intrhok}), we apply the same argument used in the proof of Theorem 3.1 in \citet[chapter 5]{Mao}: on $[0,\delta]$ we have by hypothesis a classic stochastic differential equation, and by  inequality (9.15) in Theorem 9.11 in \citet{Pascucci2011} it holds
 \begin{align}\label{estpasc1}
&\mathbb{E}[\sup_{0 \le s \le \delta}\lVert X_s \rVert^2]<\infty.
\end{align}
On the interval $[\delta,2\delta]$, we can write equation (\ref{eqXdelayed1}) as
  \begin{equation}\notag
 dX_t=\bar{b}(X_t,\xi_{t})dt+\bar{\sigma}(X_t,\xi_t)dW_t, \quad \delta \le t \le 2\delta,
 \end{equation}
where $\xi_t=X_{t-\delta}$. Once the solution on $[0,\delta]$ is known, this is again a classic SDE (without delay) with initial value $X_\delta = \xi_0$, so that by Theorem 9.11 in \citet{Pascucci2011}, there exists a constant $C_{2\delta}>0$ such that
 \begin{align}
&\mathbb{E}[\sup_{\delta \le s \le 2\delta}\lVert X_s \rVert^2]\le C_{2\delta}\left(1+\mathbb{E}[\lVert X_{\delta} \rVert^2]\right)e^{2\delta C_{2\delta}},  \label{XCdelta}
\end{align}
which is finite by (\ref{estpasc1}). Repeating this argument on the interval $[2\delta,3\delta]$, we obtain 
 \begin{align}
&\mathbb{E}[\sup_{2\delta \le s \le 3\delta}\lVert X_s \rVert^2]\le C_{3\delta}\left(1+\mathbb{E}[\lVert X_{2\delta} \rVert^2]\right)e^{3\delta C_{3\delta}} \le C_{3\delta}\left(1+\mathbb{E}[\sup_{\delta \le s \le 2\delta}\lVert X_s \rVert^2]\right)e^{3\delta C_{3\delta}}<\infty \notag
\end{align}
by \eqref{XCdelta}. Recursively we have 
\begin{align}
&\mathbb{E}[\sup_{(k-1)\delta \le s \le k\delta}\lVert X_s \rVert^2]<\infty \notag.
\end{align}
Then,
\begin{align}
\sup_{0 \le s \le t}\mathbb{E}[\lVert  X_s \rVert^2] &=  \sup_{s \in [\bar k \delta,(\bar k+1) \delta]}\mathbb{E}[\lVert  X_s \rVert^2]<\infty,
\end{align}
for some $\bar k$ with $[\bar k \delta, (\bar k +1)\delta] \subseteq [0,t]$. $ \quad \Box \notag$


 \section{Mean field limit}\label{seclimit}

We now study a mean field limit for the system of banks (\ref{rhonobubble})-(\ref{rhobubble}) for large $n$. \\
Define the processes $\tilde{\rho}^i=(\tilde{\rho}^i_t)_{t \ge 0}$, $i =1, \dots, n$, $\bar{\rho}^{k,B}=(\bar{\rho}^{k,B}_t)_{t \ge 0}$, $k=1,\dots,m$, and $\nu=(\nu_t)_{t \ge 0}$ as the solutions of the following system of SDEs for $ t \ge \delta$:
\begin{align}
&d\tr^i_t=-\lambda \tilde{\rho}_t^i dt + \sigma_1dW_{t}^i, \label{rhotilde} \\
&d\nu_t=\left( \varphi(t,t-\delta)+\frac{1}{m} \sum_{k=1}^m f^B\left(\br_{t-\delta}^{k,B}-\nu_{t-\delta}-\mathbb{E}[\tr^i_{t-\delta}]\right)\left(\br_t^{k,B}-\nu_t-\mathbb{E}[\tr^i_t]\right)+\lambda \mathbb{E}[\tr^i_t]\right)dt,\label{nu} \\
&d\br_t^{k,B}=  \left( \varphi(t,t-\delta)+\frac{1}{m-1} \sum_{\ell=1, \ell \ne k}^m f^B\left(\br_{t-\delta}^{\ell,B}-\nu_{t-\delta}-\mathbb{E}[\tr^i_{t-\delta}]\right)\left(\br_t^{\ell,B}-\nu_t-\mathbb{E}[\tr^i_t]\right)\right)dt\notag \\
&\qquad \quad +\left(\mu_t+\lambda(\mathbb{E}[\tr^i_t]+ \nu_t-\br_t^{k,B})\right)dt + \sigma_2dW_t^{k,B}+\sigma_B dB_t^1, \label{barrhobubble}
\end{align}
with
\begin{equation}\label{phi}
\varphi(t,t-\delta):=\mathbb{E}\left[f^P\left(\tr_{t-\delta}^i-\mathbb{E}[\tr_{t-\delta}^i]\right)\left(\tr_{t}^i-\mathbb{E}[\tr_t^i]\right)\right], \quad t \ge \delta.	
\end{equation}
For $t \in [0,\delta]$ we assume that $(\tr_t)_{0 \le t \le \delta}$, $(\nu_t)_{0 \le t \le \delta}$ and $(\br^{k,B}_t)_{0 \le t \le \delta}$ satisfy (\ref{rhotilde})-(\ref{barrhobubble}) for $\delta=0$, with initial conditions $\tr_0^i=\rho_0 \in \mathbb{R}$, $\nu_0=0$, $\br_0^{k,B}=\rho_0^{k,B} \in \mathbb{R}$.  \\
 Note that in equation (\ref{nu}) the expression of $\varphi$ is independent of the choice of $\tilde{\rho}^i$ since $\tilde{\rho}^i$, $i =1,\dots,n$,  are identically distributed. For the same reason, the process $\nu$ in (\ref{nu}) does not depend on $\tilde{\rho}^i$.\\
Set
\begin{equation}\label{barrhonobubble1}
\br^i:=\tr^i+\nu, \quad i = 1, \dots, n.
\end{equation}
In particular, {\small
\begin{align}\label{barrhonobubble}
\br^i_t=&\bar{\rho}_{\delta}^i+\int_{\delta}^t \left( \varphi(s,s-\delta)+\frac{1}{m} \sum_{k=1}^m f^B(\br_{s-\delta}^{k,B}-\nu_{s-\delta}-\mathbb{E}[\tr^i_{s-\delta}])
\left(\br_s^{k,B}-\nu_s-\mathbb{E}[\tr^i_s]\right)+\lambda(\mathbb{E}[\tr^i_s]-\tr^i_s)\right)ds\notag \\&+\sigma_1 W_{s}^i, \quad t \ge\delta.
\end{align}}

\begin{proposition}\label{propexlim}
Under Assumption \ref{f}, for every $\delta \ge 0$ there exists a unique strong solution of the system of SDEs (\ref{rhotilde})-(\ref{barrhobubble}). 
In particular, it holds
\begin{align}
&\sup_{0 \le s \le t}\mathbb{E}[|\nu_s|^2]<\infty, \quad  0 < t <\infty, \label{intnu}\\
&\sup_{0 \le s \le t}\mathbb{E}[|\rho^{k,B}_s|^2]<\infty, \quad  0 < t <\infty,\quad k=1,\dots,m. \label{intrhoB}
\end{align}
\end{proposition}
\textit{Proof.} For the sake of simplicity we take $\lambda=1$. It is well known that (\ref{rhotilde}) admits a unique strong solution. As before, we start by proving existence and uniqueness of the strong solution of (\ref{nu})-(\ref{barrhobubble}) when $\delta=0$. The system given by  (\ref{nu}), (\ref{barrhobubble}) and (\ref{dMu}) can be written as an $(m+2)$-dimensional SDE
\begin{equation}\label{eqX}
dX_t=b(t,X_t)dt+\sigma(t,X_t)dW_t, \quad t \ge 0,
\end{equation}
where $W=(W^{B,1}_t,\dots,W^{B,m}_t,B^1_t,B^2_t)_{t \ge 0}$, and
\begin{equation}\label{eqb}
b(t,x)=\left( \begin{array}{c}\varphi(t)+\frac{1}{m}\sum_{k=1}^m f^B(x_k-x_1-\psi(t))(x_k-x_1-\psi(t))+\psi(t),  \\
\varphi(t)+\frac{1}{m-1}\sum_{\ell=3}^{m+1} f^B(x_\ell-x_1-\psi(t))(x_\ell-x_1-\psi(t))+x_1+x_{m+2}-x_2+\psi(t),  \\
  \vdots \\
\varphi(t)+\frac{1}{m-1}\sum_{\ell=2}^{m} f^B(x_\ell-x_1-\psi(t))(x_\ell-x_1-\psi(t))+x_1+x_{m+2}-x_{m+1}+\psi(t),  \\ 
\tilde{b}(x_{m+2}) \\ \end{array}\right)
\end{equation}
with $\psi(t)=\mathbb{E}[\tilde{\rho}_t^i]$ and 
\begin{align}\label{varphi}
\varphi(t):=\mathbb{E}\left[f^P\left(\tr^i_{t}-\mathbb{E}[\tr^i_{t}]\right)\left(\tr_{t}^i-\mathbb{E}[\tr_t^i]\right)\right], \quad t \ge 0.
\end{align}
The $(m+2) \times (m+2)$ matrix $\sigma(x)$ has the form
\begin{equation}\label{matsigma}
\sigma(t,x)= \left( \begin{array}{cccccc}
0 & 0 &\dots & 0 & 0 & 0\\
\sigma_2 & 0 & \dots & 0 & \sigma_B & 0 \\
0 & \sigma_2 & \dots & 0 & \sigma_B & 0 \\
\vdots & \vdots & \ddots & 0 & \sigma_B & 0 \\
0 & 0 & \dots & \sigma_2 & \sigma_B & 0\\
0 & 0 & \dots & 0 & 0 & \tilde{\sigma}(x_{m+2})\\
 \end{array} \right).
\end{equation}
As before, we rely on Theorem 9.11 in \citet{Pascucci2011}. We have to show that $b$ and $\sigma$ defined in \eqref{eqb} and \eqref{matsigma} respectively are Lipschitz continuous in $x$ uniformly in $t$ and that for each constant $T>0$ there exists some $\tilde{C}$ such that for all $t \in [0, T]$ it holds
$$
\lVert \sigma(t,x) \rVert^2 + \lVert b(t,x) \rVert^2 \le \tilde{C} (1+ \lVert x \rVert^2).
$$
We begin by proving the first condition. The Lipschitz property clearly holds for $\sigma$, since $\tilde{\sigma}$ is Lipschitz  by hypothesis.
 Take now $x=(x_1,\dots,x_{m+2})$, $x'=(x'_1,\dots,x'_{m+2})$.  We have that
 \begin{align}
&|b_1(t,x)-b_1(t,x')| \notag \\& \quad \le  \frac{1}{m}\sum_{k=1}^m |f^B\left(x_k-x_1-\psi(t)\right)\left(x_k-x_1-\psi(t)\right)-f^B\left(x'_k-x'_1-\psi(t)\right)\left(x'_k-x'_1-\psi(t)\right)| \notag \\
& \quad \le K_2\frac{1}{m}\sum_{k=1}^m |(x_k-x_1-\psi(t))-(x'_k-x'_1-\psi(t))| \notag \\
& \quad  = K_2\left(\frac{1}{m}\sum_{k=1}^m |x_k-x'_k|+|x_1-x'_1|\right).\notag
\end{align}
Similarly, for $k=2,\dots,m+1$ we have
$$
|b_k(t,x)-b_k(t,x')| \le K_2\left(\frac{1}{m-1}\sum_{\ell=1,\ell \ne k}^m |x_\ell-x'_\ell|+|x_1-x'_1|+|x_k-x'_k|\right).
$$
With computations as in the proof of Proposition \ref{promex} we obtain that for $t \ge 0$ it holds
\begin{align}\label{lipschitz}
\lVert b(t,x)-b(t,x') \rVert^2 \le \bar{C} \lVert x-x' \rVert^2,  
\end{align}
for some appropriate $\bar{C}$.\\
We now show the second condition, i.e. that for $t \in [0,T]$ it holds
\begin{equation}\label{linear1}
\lVert \sigma(t,x) \rVert^2 + \lVert b(t,x) \rVert^2 \le \tilde{C} (1+ \lVert x \rVert^2), 
\end{equation}
for some $\tilde{C}>0$. By (\ref{matsigma}) we can focus only on $\rVert b(t,x) \rVert$. 
The computations are here the same as in Proposition \ref{promex}, but we have to estimate the term $\varphi(t)$ from (\ref{varphi}). Since
\begin{align}
|\varphi(t)|=&\left|\mathbb{E}\left[f^P\left(\tr^i_{t}-\mathbb{E}[\tr_{t}^i]\right)\left(\tr_{t}^i-\mathbb{E}[\tr^i_t]\right)\right]\right| \le  K_2 \mathbb{E}[|\tr^i_{t}-\mathbb{E}[\tr_t^i]|], \notag \\
& \le K_2 \left(\mathbb{E}[|\tr_{t}^i-\mathbb{E}[\tr^i_t]|^2]\right)^{1/2} \le K_2\left(\frac{\sigma_1^2}{2}(1-e^{-2t})\right)^{1/2} \le K_2\frac{\sigma_1}{\sqrt{2}},\notag
\end{align}
  (\ref{linear1}) follows by the proof of Proposition \ref{promex}. \\
Inequalities (\ref{intnu}) and (\ref{intrhoB}) follow since, by Theorem 9.11 in \citet{Pascucci2011}, (\ref{lipschitz}) and (\ref{linear1}) guarantee that the second moments of the solution of (\ref{eqX}) are finite.  \\
 The proof for the case $\delta>0$, based on Theorem 3.1 in \citet[chapter 5]{Mao}, is analogous to the one of Proposition \ref{promex}. 
 \quad $\Box$ \\

Denote $|x-y|_t^*=\sup_{s \le t}|x_s-y_s|$. We have the following
\begin{theorem}\label{theoremlim}
Fix $i \in \mathbb{N}$. Under Assumption \ref{f}, for any $ t \in [0,\infty)$ and $\delta \ge 0$ it holds 
$$\lim_{n \to \infty} \left(\mathbb{E}\left[|\rho^{i,n}-\br^i|^*_t\right]+\mathbb{E}[|\rho^{k,B}-\br^{k,B}|_t^*]\right)=0,  \quad k=1,\dots,m,$$
 where $\rho^{i,n}$, $\br^i$, $\rho^{k,B}$, $\br^{k,B}$ are defined in (\ref{rhonobubble}), (\ref{barrhonobubble}), (\ref{rhobubble}), (\ref{barrhobubble}) respectively. 
\end{theorem}
Before proving Theorem \ref{theoremlim}, we give the following

\begin{proposition}\label{lemmalim}
 Under Assumption \ref{f}, for $0\le\delta <\infty$, it holds
\begin{align}
& \lim_{n \to \infty}  \int_{0}^{\delta} \mathbb{E}\bigg[\Big|\frac{1}{n}\sum_{i=1}^n f^P(\bar{\rho}_{s}^{i}-\bar{A}^n_{s}) (\bar{\rho}_s^{i}-\bar{A}_s^n)- \mathbb{E}\left[f^P\left(\tr_{s}^i-\mathbb{E}[\tr_{s}^i]\right)\left(\tr_{s}^i-\mathbb{E}[\tr_s^i]\right)\right]\Big|\bigg]ds=0, \label{lemmafirst}\\
\intertext{and}
 &\lim_{n \to \infty}  \int_{\delta}^t \mathbb{E}\bigg[\Big|\frac{1}{n}\sum_{i=1}^n f^P(\bar{\rho}_{s-\delta}^{i}-\bar{A}^n_{s-\delta}) (\bar{\rho}_s^{i}-\bar{A}_s^n)-  \mathbb{E}\left[f^P\left(\tr_{s-\delta}^i-\mathbb{E}[\tr_{s-\delta}^i]\right)\left(\tr^i_{s}-\mathbb{E}[\tr^i_s]\right)\right]\Big|\bigg]ds=0,\notag
 \end{align}
 for $0 \le \delta \le t <\infty$,  where $\tilde{\rho}^i$ and $\bar{\rho}^i$ satisfy (\ref{rhotilde}) and (\ref{barrhonobubble}), respectively, and 
 \begin{equation}\label{baraverage}
  \bar{A}_t^n=\frac{1}{m+n}\left(\sum_{r=1}^n \bar{\rho}_t^{r}+\sum_{h=1}^m \bar{\rho}_t^{h,B}\right), \quad t \ge 0. 
  \end{equation}
  \end{proposition}

\textit{Proof}. We limit ourselves to prove the second limit, since the first one follows as a particular case. 
Let us write, for $t \ge \delta>0$,
\begin{align}
&\mathbb{E}\bigg[\Big|\frac{1}{n}\sum_{i=1}^n f^P(\bar{\rho}_{t-\delta}^{i}-\bar{A}^n_{t-\delta}) (\bar{\rho}_t^{i}-\bar{A}_t^n)-  \mathbb{E}\left[f^P\left(\tr^i_{t-\delta}-\mathbb{E}[\tr^i_{t-\delta}]\right)\left(\tr_{t}^i-\mathbb{E}[\tr^i_t]\right)\right]\Big|\bigg] \notag \\
&\quad \le \frac{1}{n}\sum_{i=1}^n\mathbb{E}\bigg[\Big| f^P(\bar{\rho}_{t-\delta}^{i}-\bar{A}^n_{t-\delta}) (\bar{\rho}_t^{i}-\bar{A}_t^n)-f^P(\tilde{\rho}_{t-\delta}^i-\mathbb{E}[\tilde{\rho}_{t-\delta}^i]) (\tilde{\rho}_t^i-\mathbb{E}[\tilde{\rho}_{t}^i])\Big|\bigg] \notag \\
& \quad \quad + \mathbb{E}\bigg[\Big|\frac{1}{n}\sum_{i=1}^n f^P(\tilde{\rho}_{t-\delta}^i-\mathbb{E}[\tilde{\rho}_{t-\delta}^i]) (\tilde{\rho}_t^i-\mathbb{E}[\tilde{\rho}^i_{t}])-  \mathbb{E}\left[f^P\left(\tr_{t-\delta}^i-\mathbb{E}[\tr_{t-\delta}^i]\right)\left(\tr_{t}^i-\mathbb{E}[\tr_t^i]\right)\right]\Big|\bigg],\notag
\end{align}
 since $\bar{\rho}^i$, $i=1,\dots, n$ are identically distributed and the same holds for  $\tilde{\rho}^i$, $i=1,\dots, n$.\\
By (\ref{barrhonobubble1}) we have that
\begin{align}
\bar{A}_t^n=\frac{1}{m+n}\left(\sum_{r=1}^n \bar{\rho}_t^{r}+\sum_{h=1}^m \bar{\rho}_t^{h,B}\right)=\frac{1}{m+n}\left(n\nu_t+\sum_{r=1}^n \tilde{\rho}_t^{r}+\sum_{h=1}^m \bar{\rho}_t^{h,B}\right), \notag
\end{align}
so that
\begin{align}
\lim_{n \to \infty} \bar{A}_t^n  =\nu_t+\lim_{n \to \infty}\frac{1}{m+n}\sum_{r=1}^n \tilde{\rho}_t^{r}=
\nu_t+\mathbb{E}[\tilde{\rho}^i_t], \quad a.s.,\notag
\end{align}
 by (\ref{intrhok}) and the law of large numbers, as $\tr^i$, $i = 1, \dots, n$, are independent and identically distributed. 
Then we have
\begin{align}
\lim_{n \to \infty}f^P(\bar{\rho}_{t-\delta}^{i}-\bar{A}^n_{t-\delta}) (\bar{\rho}_t^{i}-\bar{A}_t^n)=&f^P\left(\nu_{t-\delta}+\tr^i_{t-\delta}-(\nu_{t-\delta}+\mathbb{E}[\tilde{\rho}^i_{t-\delta}])\right) \left(\nu_{t}+\tr^i_{t}-(\nu_{t}+\mathbb{E}[\tilde{\rho}^i_{t}])\right)  \notag \\
&=f^P\left(\tilde{\rho}_{t-\delta}^i-\mathbb{E}[\tilde{\rho}_{t-\delta}^i]\right) \left(\tilde{\rho}_t^i-\mathbb{E}[\tilde{\rho}^i_{t}]\right) \quad a.s.\label{limit}
\end{align}
%
  We now prove that the family of random variables $\{\frac{1}{n}\sum_{i=1}^n  f^P(\bar{\rho}_{s-\delta}^{i}-\bar{A}^n_{s-\delta}) (\bar{\rho}_s^{i}-\bar{A}_s^n)\}_{n \in \mathbb{N}}$ is uniformly integrable for every $s \in [\delta,t]$, so that convergence almost surely implies convergence in $L^1$. \\
By point (iii) of Theorem 11 in \citet[chapter 1]{ProtterBook} it is enough to prove that for every $s \in [\delta,t]$,
\begin{equation}\label{unifint}
\sup_n \mathbb{E}\left[\left(\frac{1}{n}\sum_{i=1}^n  f^P(\bar{\rho}_{s-\delta}^{i}-\bar{A}^n_{s-\delta}) (\bar{\rho}_s^{i}-\bar{A}_s^n)\right)^2\right]<\infty.
\end{equation}
For every $s \in [\delta,t]$, we have that
\begin{align}
&\mathbb{E}\bigg[\Big(\frac{1}{n}\sum_{i=1}^n   f^P(\bar{\rho}_{s-\delta}^{i}-\bar{A}^n_{s-\delta}) (\bar{\rho}_s^{i}-\bar{A}_s^n)\Big)^2\bigg]\le (K_2)^2 \mathbb{E}\bigg[\Big(\frac{1}{n}\sum_{i=1}^n | \bar{\rho}_s^{i}-\bar{A}_s^n|\Big)^2\bigg] \notag \\ & \quad \le (K_2)^2 \mathbb{E}\bigg[\Big(  (1-n/(m+n))|\nu_s|+\left|\tilde{\rho}_s^i\right|+\frac{1}{m+n}\sum_{r=1}^n\left|\tilde{\rho}_s^{r}\right|+\frac{1}{m+n}\sum_{h=1}^m\left|\bar{\rho}_s^{h,B}\right|\Big)^2\bigg] \notag \\
& \quad \le (K_2)^2 \mathbb{E}\bigg[\Big(|\nu_s|+\left|\tilde{\rho}_s^i\right|+\frac{1}{n}\sum_{r=1}^n\left|\tilde{\rho}_s^{r}\right|+\frac{1}{m}\sum_{h=1}^m\left|\bar{\rho}_s^{h,B}\right|\Big)^2\bigg] \notag \\
& \quad \le 4(K_2)^2 \bigg(\mathbb{E}\Big[|\nu_s|^2+|\tilde{\rho}_s^i|^2+\sum_{k=1}^m|\bar{\rho}_s^{k,B}|^2\Big]+\mathbb{E}\Big[\Big(\frac{1}{n}\sum_{r=1}^n\left|\tilde{\rho}_s^{r}\right|\Big)^2\Big] \bigg) \notag \\
&\quad  \le 4(K_2)^2 \bigg(\mathbb{E}\Big[|\nu_s|^2+|\tilde{\rho}_s^i|^2+\sum_{k=1}^m|\bar{\rho}_s^{k,B}|^2\Big]+\frac{1}{n}\mathbb{E}\bigg[\sum_{r=1}^n |\tr_s^r|^2\bigg] \bigg). \notag \\
&  \quad\le 4(K_2)^2 \bigg(\mathbb{E}\Big[|\nu_s|^2+|\tilde{\rho}_s^i|^2+\sum_{k=1}^m|\bar{\rho}_s^{k,B}|^2\Big]+\mathbb{E}[ |\tr_s^i|^2] \bigg)<\infty \notag ,
\end{align}
by (\ref{intnu}) and (\ref{intrhoB}) and because $\mathbb{E}|\tilde{\rho}_s^i|^2]<\infty$. 
Hence,  $\{\frac{1}{n}\sum_{i=1}^n  f^P(\bar{\rho}_{s-\delta}^{i}-\bar{A}^n_{s-\delta}) (\bar{\rho}_s^{i}-\bar{A}_s^n)\}_{n \in \mathbb{N}}$ is uniformly integrable and we obtain therefore by (\ref{limit}) that
\begin{equation}\notag
\lim_{n\to \infty} \mathbb{E}\bigg[\Big| f^P(\bar{\rho}_{t-\delta}^{i}-\bar{A}^n_{t-\delta}) (\bar{\rho}_t^{i}-\bar{A}_t^n)-f^P(\tilde{\rho}_{t-\delta}^i-\mathbb{E}[\tilde{\rho}_{t-\delta}^i]) (\tilde{\rho}_t^i-\mathbb{E}[\tilde{\rho}_{t}^i])\Big|\bigg]=0. 
\end{equation}
Moreover, for $\delta \le s \le t$ it holds
\begin{align}
&\mathbb{E}\bigg[\Big| f^P(\bar{\rho}_{t-\delta}^{i}-\bar{A}^n_{t-\delta}) (\bar{\rho}_t^{i}-\bar{A}_t^n)-f^P(\tilde{\rho}_{t-\delta}^i-\mathbb{E}[\tilde{\rho}_{t-\delta}^i]) (\tilde{\rho}_t^i-\mathbb{E}[\tilde{\rho}_{t}^i])\Big|\bigg] \notag \\ & \quad \le 
K_1 (\mathbb{E}[|\bar{\rho}_t^{i}-\bar{A}_t^n|] +\mathbb{E}[|\tilde{\rho}_t^i-\mathbb{E}[\tilde{\rho}_{t}^i|]), \notag
\end{align}
where the second term belongs to $L^1\left([\delta,t]\right)$  and does not depend on $n$. On the other hand,
we have \begin{align}
 \int_0^t \mathbb{E}[|\bar{\rho}_s^{i}-\bar{A}_s^n|]ds&\le\int_0^t \mathbb{E}\left[|\tilde{\rho}_s^{i}|+\left(1-n/(m+n)\right)|\nu_s|+\frac{1}{m+n}\sum_{r=1}^{n}|\tilde{\rho}_s^r|+\frac{1}{m+n}\sum_{h=1}^{m}|\bar{\rho}_s^{h,B}|\right]ds \notag \\
 & \le \int_0^t \mathbb{E}\left[2|\tilde{\rho}_s^{i}|+|\nu_s|+|\bar{\rho}^{h,B}_s|\right]ds \notag \\
 & \le t \sup_{0 \le s \le t}\mathbb{E}\left[2|\tilde{\rho}_s^{i}|+|\nu_s|+|\bar{\rho}_s^{h,B}|\right]<\infty \label{footnote},
 \end{align}by (\ref{intnu}) and (\ref{intrhoB}). We can then apply the dominated convergence theorem to obtain, for $t \in [\delta,\infty)$,
\begin{equation}\label{limitprooflemma}
\lim_{n \to \infty}  \int_{\delta}^t \mathbb{E}\bigg[\Big| f^P(\bar{\rho}_{s-\delta}^{i}-\bar{A}^n_{s-\delta}) (\bar{\rho}_s^{i}-\bar{A}_s^n)-f^P(\tilde{\rho}_{s-\delta}^i-\mathbb{E}[\tilde{\rho}_{s-\delta}^i]) (\tilde{\rho}_s^i-\mathbb{E}[\tilde{\rho}_{s}^i])\Big|\bigg] ds =0, \quad t \ge \delta.
\end{equation}
It remains to show that for $t \ge \delta$ it holds
\begin{equation}\label{lastlimit}
\lim_{n \to \infty}  \int_{\delta}^t \mathbb{E}\bigg[\Big|\frac{1}{n}\sum_{i=1}^n f^P(\tilde{\rho}_{s-\delta}^i-\mathbb{E}[\tilde{\rho}_{s-\delta}^i]) (\tilde{\rho}_s^i-\mathbb{E}[\tilde{\rho}^i_{s}])-  \mathbb{E}\left[f^P\left(\tr_{s-\delta}^i-\mathbb{E}[\tr_{s-\delta}^i]\right)\left(\tr_{s}^i-\mathbb{E}[\tr_s^i]\right)\right]\Big|\bigg]ds=0.
\end{equation}
Since $\tilde{\rho}^i$, $i=1,\dots, n$, are independent and identically distributed, we have that, $\text{for } \delta \le s \le t$,
$$
\lim_{n \to \infty}  \mathbb{E}\bigg[\Big|\frac{1}{n}\sum_{i=1}^n f^P(\tilde{\rho}_{s-\delta}^i-\mathbb{E}[\tilde{\rho}_{s-\delta}^i]) (\tilde{\rho}_s^i-\mathbb{E}[\tilde{\rho}^i_{s}])-  \mathbb{E}\left[f^P\left(\tr_{s-\delta}^i-\mathbb{E}[\tr_{s-\delta}^i]\right)\left(\tr_{s}^i-\mathbb{E}[\tr_s^i]\right)\right]\Big|\bigg]=0.
$$
Then limit \eqref{lastlimit} follows by the dominated convergence theorem, by Assumption \ref{f} and since the Ornstein-Uhlenbeck process has finite moments, see the computations in \eqref{footnote}. $\quad \Box$\\

\textit{Proof of Theorem \ref{theoremlim}}.  
We suppose by simplicity $\lambda=1$ and we proceed by steps, starting from the case when $0\le t< \delta$, i.e. when there is no delay in equations (\ref{rhonobubble})-(\ref{rhobubble}) and (\ref{nu})-(\ref{barrhobubble}).\\
\textit{First step: case $0\le t< \delta$.} \\
For every $i = 1, \dots,n$ and $t \in [0,\delta)$, we have
\begin{align}
\r^{i,n}_t-\br_t^i&=\int_0^t \Delta_s^nds, \notag
\end{align}
where 
\begin{align}\notag
\Delta_s^n=&\frac{1}{n-1}\sum_{j=1,j \ne i}^n  f^P(\r_s^{j,n}-A_s^n)(\r^{j,n}_s-A^n_s)-\mathbb{E}\left[f^P\left(\tr_{s}^i-\mathbb{E}[\tr_{s}^i]\right)\left(\tr_{s}^i-\mathbb{E}[\tr_s^i]\right)\right]\notag \\&+\frac{1}{m}\sum_{k=1}^m  \left(f^B(\r_s^{k,B}-A_s^n)(\r^{k,B}_s-A_s^n)-f^B(\br_s^{k,B}-\nu_{s}^n-\mathbb{E}[\tilde{\rho}^i_{s}])(\br^{k,B}_s-\nu_s-\mathbb{E}[\tilde{\rho}^i_s])\right) \notag \\
 &\quad -(\r_s^{i,n}-\br^i_s) +(A_s^n-\bar{A}^n_s)+(\bar{A}_s^n-\mathbb{E}[\tilde{\rho}^i_s]-\nu_s) \notag.
\end{align}
Thus
\begin{align}
 |\rho^{i,n}-\br^i|_t^* =& \sup_{s \le t} \left|\int_0^s \Delta_u^n du\right|\le \sup_{s \le t} \int_0^s \left|\Delta_u^n\right|du\notag = \int_0^t \left|\Delta_u^n \right|du.\notag 
\end{align}

Therefore, for every $i = 1, \dots,n$ and $t \ge 0$, we have
\begin{align}\label{1diff}
&\mathbb{E}[|\r^{i,n}-\br^i|^*_t]\le \mathbb{E}\left[\int_0^t |\Delta_s^n|ds\right] \notag \\
& \le \int_0^t \mathbb{E}\bigg[\Big|\frac{1}{n-1}\sum_{j=1,j \ne i}^n  (f^P(\r_s^{j,n}-A_s^n)(\r^{j,n}_s-A_s^n) - f^P(\bar{\r}_s^{j}-\bar{A}_s^n)(\bar{\r}^{j}_s-\bar{A}_s^n))\Big|\bigg]ds\notag \\
&  \quad + \int_{0}^t \mathbb{E}\bigg[\Big|\frac{1}{n-1}\sum_{j=1,j\ne i}^n f^P(\bar{\rho}_{s}^{j}-\bar{A}_{s}^n) (\bar{\rho}_s^{j}-\bar{A}_s^n)-  \mathbb{E}\left[f^P\left(\tr_{s}^i-\mathbb{E}[\tr_{s}^i]\right)\left(\tr^i_{s}-\mathbb{E}[\tr^i_s]\right)\right]\Big|\bigg]ds \notag \\
&  \quad \quad + \int_0^t\mathbb{E}\bigg[\Big| \frac{1}{m}\sum_{k=1}^m  \left(f^B(\r_s^{k,B}-A_s^n)(\r^{k,B}_s-A_s^n)-f^B(\br_s^{k,B}-\bar{A}_s^n)(\br^{k,B}_s-\bar{A}_s^n)\right)\Big|\bigg]ds \notag \\
& \qquad \quad + \int_{0}^t \mathbb{E}\bigg[\frac{1}{m}\sum_{k=1}^m  \left|f^B(\bar{\r}_s^{k,B}-\bar{A}_s^n)(\bar{\r}^{k,B}_s-\bar{A}_s^n)-f^B(\br_s^{k,B}-\nu_{s}^n-\mathbb{E}[\tilde{\rho}^i_{s}])(\br^{k,B}_s-\nu_s-\mathbb{E}[\tilde{\rho}^i_s])\right|\bigg]ds \notag \\
 & \qquad \quad \quad +  \int_0^t \mathbb{E}[|\r_s^{i,n}-\br^i_s|]ds+ \int_0^t \mathbb{E}[|A_s^{n}-\bar{A}^n_s|]ds+\int_0^t\mathbb{E}\left[|\bar{A}_s^n-\mathbb{E}[\tilde{\rho}^i_s]-\nu_s|\right]ds.
\end{align}
By (\ref{eqf}) it holds 
\begin{align}\label{1int1}
&\int_0^t \mathbb{E}\bigg[\Big|\frac{1}{n-1}\sum_{j=1,j \ne i}^n  (f^P(\r_s^{j,n}-A_s^n)(\r^{j,n}_s-A_s^n) - f^P(\bar{\r}_s^{j}-\bar{A}_s^n)(\bar{\r}^{j}_s-\bar{A}_s^n))\Big|\bigg]ds \notag \\
 & \quad \le \frac{1}{n-1}\sum_{j=1,j \ne i}^n \int_0^t  \mathbb{E}\bigg[\Big| f^P(\r_s^{j,n}-A_s^n)(\r^{j,n}_s-A_s^n) - f^P(\bar{\r}_s^{j}-\bar{A}_s^n)(\bar{\r}^{j}_s-\bar{A}_s^n)\Big|\bigg]ds \notag\\
 & \quad \le K_2\frac{1}{n-1}\sum_{j=1,j \ne i}^n \int_0^t  \mathbb{E}\bigg[\Big| (\r^{j,n}_s-A_s^n) - (\bar{\r}^{j}_s-\bar{A}_s^n)\Big|\bigg]ds \notag\\ 
  &  \quad \le K_2\frac{1}{n-1} \sum_{j=1,j \ne i}^n \int_0^t  \mathbb{E}\left[\left|\r^{j,n}_s - \br_s^j\right|+\left|A^n_s-\bar{A}^n_s\right|\right]ds \notag \\
  & \quad = K_2 \int_0^t  \mathbb{E}\left[\left|\r^{i,n}_s - \br_s^i\right|\right]ds+K_2 \int_0^t  \mathbb{E}\left[\left|A^{n}_s - \bar{A}_s^n\right|\right]ds, \quad t \ge 0.    
\end{align}
By (\ref{average}) and (\ref{baraverage}) we have that
\begin{align}
  &  \int_0^t  \mathbb{E}\left[\left|A^{n}_s - \bar{A}_s^n\right|\right]ds \le 
     \int_0^t \mathbb{E}\Big[\frac{1}{m+n}\sum_{r=1}^n\left|\rho^{r,n}_s - \bar{\rho}_s^r\right|\Big]ds+ \int_0^t \mathbb{E}\Big[  \frac{1}{m+n} \sum_{k=1}^m\left|\rho^{h,B}_s - \bar{\rho}_s^{h,B}\right|\Big]ds\notag \\
&   \le \int_0^t \mathbb{E}\left[\left|\rho^{i,n}_s - \bar{\rho}_s^i\right|\right]ds+\int_0^t\mathbb{E}\left[ \left|\rho^{k,B}_s - \bar{\rho}_s^{k,B}\right|\right]ds, \quad t \ge 0,    \label{A}
\end{align}
because all $\rho^i$, $i=1,\dots,n$, and $\rho^{k,B}$, $k=1,\dots,m$,  are  identically distributed, respectively.\\
We can conclude by (\ref{1int1}) and (\ref{A}) that
\begin{align}\label{1intt1}
&\int_0^t \mathbb{E}\bigg[\Big|\frac{1}{n-1}\sum_{j=1,j \ne i}^n  (f^P(\r_s^{j,n}-A_s^n)(\r^{j,n}_s-A_s^n) - f^P(\bar{\r}_s^{j}-\bar{A}_s^n)(\bar{\r}^{j}_s-\bar{A}_s^n))\Big|\bigg]ds \notag \\
& \quad \le 2 K_2 \int_0^t  \mathbb{E}\left[\left|\r^{i,n}_s - \br_s^i\right|\right]ds+K_2\int_0^t\mathbb{E}\left[ \left|\rho^{k,B}_s - \bar{\rho}_s^{k,B}\right|\right]ds \notag \\
&  \quad \le 2 K_2 \int_0^t  \mathbb{E}\left[\left|\r^{i,n} - \br^i\right|_s^*\right]ds+K_2\int_0^t\mathbb{E}\left[ \left|\rho^{k,B} - \bar{\rho}^{k,B}\right|^*_s\right]ds,  \quad t \ge 0. 
\end{align}
Similarly,
\begin{align}\label{1int2}
& \int_0^t\mathbb{E}\bigg[\Big| \frac{1}{m}\sum_{k=1}^m  \left(f^B(\r_s^{k,B}-A_s^n)(\r^{k,B}_s-A_s^n)-f^B(\br_s^{k,B}-\bar{A}_s^n)(\br^{k,B}_s-\bar{A}_s^n)\right)\Big|\bigg]ds \notag \\
& \quad  \le K_2 \int_0^t  \mathbb{E}\left[\left|\r^{i,n} - \br^i\right|_s^*\right]ds+2K_2 \int_0^t \mathbb{E}\left[\left|\r^{k,B} - \bar{\rho}^{k,B}\right|_s^*\right]ds \quad t \ge 0. 
\end{align}
From (\ref{1diff}), (\ref{A}), (\ref{1intt1}) and (\ref{1int2}) we have that for $t \ge 0$ it holds
\begin{align}
& \mathbb{E}[|\r^{i,n}-\br^i|_t^*] \notag \\ 
& \le  (3K_2+2) \int_0^t  \mathbb{E}\left[\left|\r^{i,n} - \br^i\right|^*_s\right]ds  +(3K_2+1)\int_0^t  \mathbb{E}\left[\left|\r^{k,B} - \br^{k,B}\right|_s^*\right]ds \notag \\
&  \quad + \int_{0}^t \mathbb{E}\bigg[\left|f^B(\bar{\r}_s^{k,B}-\bar{A}_s^n)(\bar{\r}^{k,B}_s-\bar{A}_s^n)-f^B(\br_s^{k,B}-\nu_{s}^n-\mathbb{E}[\tilde{\rho}^i_{s}])(\br^{k,B}_s-\nu_s-\mathbb{E}[\tilde{\rho}^i_s])\right|\bigg]ds \notag \\
& \qquad + \int_{0}^t \mathbb{E}\bigg[\Big|\frac{1}{n-1}\sum_{j=1,j\ne i}^n f^P(\bar{\rho}_{s}^{j}-\bar{A}_{s}^n) (\bar{\rho}_s^{j}-\bar{A}_s^n)-  \mathbb{E}\left[f^P\left(\tr_{s}^i-\mathbb{E}[\tr_{s}^i]\right)\left(\tr^i_{s}-\mathbb{E}[\tr^i_s]\right)\right]\Big|\bigg]ds \notag \\ 
& \qquad \quad + \int_0^t\mathbb{E}\left[|\bar{A}_s^n-\mathbb{E}[\tilde{\rho}^i_s]-\nu_s|\right]ds, \quad t \ge 0.  \label{1eqi} 
\end{align}
Proceeding as before, we find
\begin{align}
&\mathbb{E}[|\r^{k,B}-\br^{k,B}|^*_t] \notag \\ & \le (3K_2+1) \int_0^t  \mathbb{E}\left[\left|\r^{i,n} - \br^i\right|^*_s\right]ds  +(3K_2+2)\int_0^t  \mathbb{E}\left[\left|\r^{k,B} - \br^{k,B}\right|_s^*\right]ds \notag \\
& \quad  + \int_{0}^t \mathbb{E}\bigg[ \left|f^B(\bar{\r}_s^{k,B}-\bar{A}_s^n)(\bar{\r}^{k,B}_s-\bar{A}_s^n)-f^B(\br_s^{k,B}-\nu_{s}^n-\mathbb{E}[\tilde{\rho}^i_{s}])(\br^{k,B}_s-\nu_s-\mathbb{E}[\tilde{\rho}^i_s])\right|\bigg]ds \notag \\
& \qquad + \int_{0}^t \mathbb{E}\bigg[\Big|\frac{1}{n}\sum_{i=1}^n f^P(\bar{\rho}_{s}^{i}-\bar{A}_{s}^n) (\bar{\rho}_s^{i}-\bar{A}_s^n)-  \mathbb{E}\left[f^P\left(\tr_{s}^i-\mathbb{E}[\tr_{s}^i]\right)\left(\tr^i_{s}-\mathbb{E}[\tr^i_s]\right)\right]\Big|\bigg]ds \notag \\
&  \qquad \quad+ \int_0^t \mathbb{E}\left[|\bar{A}_s^{n}-\nu_s-\mathbb{E}[\tilde{\nu}^i_s|]\right]ds,   \label{1eqk} 
\end{align}
so that, summing up (\ref{1eqi}) and (\ref{1eqk}), we have 
\begin{align}
&\mathbb{E}[|\r^{i,n}-\br^i|^*_t]+\mathbb{E}[|\r^{k,B}-\br^{k,B}|_t^*] \notag \\
&\le(6K_2+3) \int_0^t  \mathbb{E}\left[\left|\r^{i,n} - \br^i\right|^*_s\right]ds  +(6K_2+3)\int_0^t  \mathbb{E}\left[\left|\r^{k,B} - \br^{k,B}\right|_s^*\right]ds \notag \\
& \quad  +2 \int_{0}^t \mathbb{E}\bigg[\left|f^B(\bar{\r}_s^{k,B}-\bar{A}_s^n)(\bar{\r}^{k,B}_s-\bar{A}_s^n)-f^B(\br_s^{k,B}-\nu_{s}^n-\mathbb{E}[\tilde{\rho}^i_{s}])(\br^{k,B}_s-\nu_s-\mathbb{E}[\tilde{\rho}^i_s])\right|\bigg]ds \notag \\
& \quad \quad+ \int_{0}^t \mathbb{E}\bigg[\Big|\frac{1}{n-1}\sum_{j=1,j\ne i}^n f^P(\bar{\rho}_{s}^{j}-\bar{A}_{s}^n) (\bar{\rho}_s^{j}-\bar{A}_s^n)-  \mathbb{E}\left[f^P\left(\tr_{s}^i-\mathbb{E}[\tr_{s}^i]\right)\left(\tr^i_{s}-\mathbb{E}[\tr^i_s]\right)\right]\Big|\bigg]ds \notag \\
& \quad  \quad\quad+ \int_{0}^t \mathbb{E}\bigg[\Big|\frac{1}{n}\sum_{i=1}^n f^P(\bar{\rho}_{s}^{i}-\bar{A}_{s}^n) (\bar{\rho}_s^{i}-\bar{A}_s^n)-  \mathbb{E}\left[f^P\left(\tr_{s}^i-\mathbb{E}[\tr_{s}^i]\right)\left(\tr^i_{s}-\mathbb{E}[\tr^i_s]\right)\right]\Big|\bigg]ds \notag \\
&  \quad\quad\quad\quad+2 \int_0^t \mathbb{E}\left[|\bar{A}_s^{n}-\nu_s-\mathbb{E}[\tilde{\nu}^i_s|]\right]ds, \quad t \ge 0.
\end{align}
We can now apply Gronwall's Lemma and obtain 
\begin{align}
&\mathbb{E}[|\r^{i,n}-\br^i|_t^*]+\mathbb{E}[|\r_t^{k,B}-\br^{k,B}_t|_s^*]  \notag \\ &\le  e^{(6K_2+3)t} \int_0^t \mathbb{E}\bigg[\Big|\frac{1}{n-1}\sum_{j=1,j\ne i}^n f^P(\bar{\rho}_{s}^{j}-\bar{A}_{s}^n) (\bar{\rho}_s^{j}-\bar{A}_s^n)-  \mathbb{E}\left[f^P\left(\tr_{s}^i-\mathbb{E}[\tr_{s}^i]\right)\left(\tr^i_{s}-\mathbb{E}[\tr^i_s]\right)\right]\Big|\bigg]ds \notag \\
& \quad + e^{(6K_2+3)t} \int_0^t \mathbb{E}\bigg[\Big|\frac{1}{n}\sum_{i=1}^n f^P(\bar{\rho}_{s}^{i}-\bar{A}_{s}^n) (\bar{\rho}_s^{i}-\bar{A}_s^n)-  \mathbb{E}\left[f^P\left(\tr_{s}^i-\mathbb{E}[\tr_{s}^i]\right)\left(\tr^i_{s}-\mathbb{E}[\tr^i_s]\right)\right]\Big|\bigg]ds \notag \\
& \quad  \quad+2e^{(6K_2+3)t} \int_{0}^t \mathbb{E}\bigg[\left|f^B(\bar{\r}_s^{k,B}-\bar{A}_s^n)(\bar{\r}^{k,B}_s-\bar{A}_s^n)-f^B(\br_s^{k,B}-\nu_{s}-\mathbb{E}[\tilde{\rho}^i_{s}])(\br^{k,B}_s-\nu_s-\mathbb{E}[\tilde{\rho}^i_s])\right|\bigg]ds \notag \\
& \quad\quad\quad+2e^{(6K_2+3)t} \int_0^t \mathbb{E}\left[|\bar{A}_s^{n}-\nu_s-\mathbb{E}[\tilde{\nu}^i_s|]\right]ds, \quad t \ge 0.
\end{align}
We can write
\begin{align}
& \int_0^t \mathbb{E}\bigg[\Big|\frac{1}{n-1}\sum_{j=1,j\ne i}^n f^P(\bar{\rho}_{s}^{j}-\bar{A}_{s}^n) (\bar{\rho}_s^{j}-\bar{A}_s^n)-  \mathbb{E}\left[f^P\left(\tr_{s}^i-\mathbb{E}[\tr_{s}^i]\right)\left(\tr^i_{s}-\mathbb{E}[\tr^i_s]\right)\right]\Big|\bigg]ds \notag \\
 & \quad \le \left(\frac{1}{n-1}-\frac{1}{n}\right) \int_0^t \mathbb{E}\bigg[\Big| \sum_{j=1,j \ne i}^nf^P(\bar{\rho}_{s}^{j}-\bar{A}_{s}^n) (\bar{\rho}_s^{j}-\bar{A}_s^n)\Big|\bigg]ds \notag \\
 & \qquad +  \int_{0}^t \mathbb{E}\bigg[\Big|\frac{1}{n}\sum_{i=1}^n f^P(\bar{\rho}_{s}^{i}-\bar{A}_{s}^n) (\bar{\rho}_s^{i}-\bar{A}_s^n)-  \mathbb{E}\left[f^P\left(\tr_{s}^i-\mathbb{E}[\tr_{s}^i]\right)\left(\tr^i_{s}-\mathbb{E}[\tr^i_s]\right)\right]\Big|\bigg]ds \notag \\
& \quad \quad \quad + \frac{1}{n} \int_0^t \mathbb{E}\left[f^P\left(\tr_{s}^i-\mathbb{E}[\tr_{s}^i]\right)\left(\tr^i_{s}-\mathbb{E}[\tr^i_s]\right)\right]ds  \quad t \ge 0, \notag
\end{align}
with
\begin{align}
 &\left(\frac{1}{n-1}-\frac{1}{n}\right) \int_0^t \mathbb{E}\bigg[\Big| \sum_{j=1,j \ne i}^n  f^P(\bar{\rho}_{s}^{i}-\bar{A}_{s}^n) (\bar{\rho}_s^{i}-\bar{A}_s^n)\Big|\bigg]ds \notag \\ & \quad \le  \frac{1}{n(n-1)} \int_0^t  \sum_{j=1,j \ne i}^n \mathbb{E}[ |f^P(\bar{\rho}_{s}^{i}-\bar{A}_{s}^n) (\bar{\rho}_s^{i}-\bar{A}_s^n)|]ds \notag \\
 & \qquad = \frac{1}{n} \int_0^t\mathbb{E}[ |f^P(\bar{\rho}_{s}^{i}-\bar{A}_{s}^n) (\bar{\rho}_s^{i}-\bar{A}_s^n)|]ds \le  \frac{K_2}{n} \int_0^t \mathbb{E}[|\bar{\rho}_s^{i}-\bar{A}_s^n|]ds, \quad t \ge 0,\notag
\end{align}

 where the last term tends to zero when $n \to \infty$  by (\ref{footnote}). 
\\
Since it can be shown, for $ t \ge 0$,  that
\begin{align}
 \lim_{n \to \infty} \int_{0}^t \mathbb{E}\bigg[\left|f^B(\bar{\r}_s^{k,B}-\bar{A}_s^n)(\bar{\r}^{k,B}_s-\bar{A}^n_s)-f^B(\br_s^{k,B}-\nu_{s}-\mathbb{E}[\tilde{\rho}^i_{s}])(\br^{k,B}_s-\nu_s-\mathbb{E}[\tilde{\rho}^i_s])\right|\bigg]ds = 0,  \notag 
 \end{align}
 and
 $$
\lim_{n \to \infty} \int_0^t \mathbb{E}\left[|\bar{A}_s^{n}-\nu_s-\mathbb{E}[\tilde{\nu}^i_s|]\right]ds=0, \quad t \ge 0,
$$ 
with the same proof as for (\ref{limitprooflemma}), then by (\ref{lemmafirst}) we obtain the result for $t \in [0,\delta)$.\\
 
 \textit{Second step: case $t \in [\delta,2\delta)$.}\\
For every $i = 1, \dots,n$ and $t \ge \delta$, we have
\begin{align}
|\r^{i,n}_t-\br_t^i|&\le \left| \int_0^{\delta} (\r^{i,n}_s-\br_s^i)ds+ \int_{\delta}^t \Delta_s^{\delta,n}ds\right|, \notag
\end{align}
where 
\begin{align}\notag
\Delta_s^{\delta,n}=&\frac{1}{n-1}\sum_{j=1,j \ne i}^n  f^P(\r_{s-\delta}^{j,n}-A_{s-\delta}^n)(\r^{j,n}_s-A^n_s)-\mathbb{E}\left[f^P\left(\tr_{t-\delta}^i-\mathbb{E}[\tr_{t-\delta}^i]\right)\left(\tr_{t}^i-\mathbb{E}[\tr_t^i]\right)\right]\notag \\&+\frac{1}{m}\sum_{k=1}^m  \left(f^B(\r_{s-\delta}^{k,B}-A_{s-\delta}^n)(\r^{k,B}_s-A_s^n)-f^B(\br_{s-\delta}^{k,B}-\nu_{s-\delta}-\mathbb{E}[\tilde{\rho}^i_{s-\delta}])(\br^{k,B}_s-\nu_s-\mathbb{E}[\tilde{\rho}^i_s])\right) \notag \\
 & -(\r_{s}^{i,n}-\br^i_{s}) +(A^n_{s}-\bar{A}_s^n)+(\bar{A}_s^n-\mathbb{E}[\tilde{\rho}^i_s]-\nu_s|) \notag.
\end{align}
Thus
\begin{align}
 |\rho^{i,n}-\br^i|_t^* &= \sup_{s \le t} \left|\int_{0}^{\delta}(\r^{i,n}_u-\br_u^i)du+\int_{\delta}^s \Delta_u^{\delta,n} du\right|\le \int_{0}^{\delta}|\r^{i,n}_u-\br_u^i|du+ \sup_{\delta \le s \le t} \int_{\delta}^s \left|\Delta_u^{\delta,n}\right|du\notag \\ &= 
\int_{0}^{\delta}|\r^{i,n}_u-\br_u^i|du+ \int_{\delta}^t \left|\Delta_u^{\delta,n}\right|du, \quad  \delta \le t . \label{ledelte}
\end{align}

For every $i = 1, \dots,n$, we have
\begin{align}\label{diff}
& \mathbb{E}\left[\int_{\delta}^t |\Delta_s^{\delta,n}|ds\right] \notag \\ & \le \int_{\delta}^t \mathbb{E}\bigg[\Big|\frac{1}{n-1}\sum_{j=1,j \ne i}^n  (f^P(\r_{s-\delta}^{j,n}-A_{s-\delta}^n)(\r^{j,n}_s-A_s^n) - f^P(\bar{\r}_{s-\delta}^{j}-\bar{A}_{s-\delta}^n)(\bar{\r}^{j}_s-\bar{A}_s^n))\Big|\bigg]ds\notag \\
&  \quad + \int_{\delta}^t \mathbb{E}\bigg[\Big|\frac{1}{n-1}\sum_{j=1,j\ne i}^n f^P(\bar{\rho}_{s-\delta}^{j}-\bar{A}_{s-\delta}^n) (\bar{\rho}_s^{j}-\bar{A}_s^n)-  \mathbb{E}\left[f^P\left(\tr_{s-\delta}^i-\mathbb{E}[\tr_{s-\delta}^i]\right)\left(\tr^i_{s}-\mathbb{E}[\tr^i_s]\right)\right]\Big|\bigg]ds \notag \\
& \quad + \int_{\delta}^t\mathbb{E}\bigg[\Big| \frac{1}{m}\sum_{k=1}^m  \left(f^B(\r_{s-\delta}^{k,B}-A_{s-\delta}^n)(\r^{k,B}_s-A_s^n)-f^B(\br_{s-\delta}^{k,B}-\bar{A}_{s-\delta}^n)(\br^{k,B}_s-\bar{A}_s^n)\right)\Big|\bigg]ds \notag \\
& \quad + \int_{\delta}^t \mathbb{E}\bigg[\frac{1}{m}\sum_{k=1}^m  \left|f^B(\bar{\r}_{s-\delta}^{k,B}-\bar{A}_{s-\delta}^n)(\bar{\r}^{k,B}_s-\bar{A}_s^n)-f^B(\br_{s-\delta}^{k,B}-\nu_{s-\delta}-\mathbb{E}[\tilde{\rho}^i_{s-\delta}])(\br^{k,B}_s-\nu_s-\mathbb{E}[\tilde{\rho}^i_s])\right|\bigg]ds \notag \\
 & \quad +  \int_{\delta}^t \mathbb{E}[|\r_s^{i,n}-\br^i_s|]ds+ \int_0^t \mathbb{E}[|A_s^{n}-\bar{A}^n_s|]ds+\int_0^t\mathbb{E}\left[|\bar{A}_s^n-\mathbb{E}[\tilde{\rho}^i_s]-\nu_s|\right]ds
, \quad  \delta \le t.
\end{align}
By (\ref{eqf}) it holds 
\begin{align}\label{int11}
& \int_{\delta}^t \mathbb{E}\bigg[\Big|\frac{1}{n-1} \sum_{j=1,j \ne i}^n (f^P(\r_{s-\delta}^{j,n}-A^n_{s-\delta})(\r^{j,n}_s-A^n_{s}) - f^P(\br^j_{s-\delta}-\bar{A}^n_{s-\delta})(\br_s^j-\bar{A}_s^n))\Big|\bigg]ds  \notag \\ 
& \quad \le \frac{1}{n-1}\sum_{j=1,j \ne i}^n \int_{\delta}^t  \mathbb{E}\bigg[\Big| f^P(\r_{s-\delta}^{j,n}-A^n_{s-\delta})\left((\r^{j,n}_s-A^n_{s})+(\br_s^j-\bar{A}^n_{s})\right)\Big|\bigg]ds \notag \\
&\qquad + \frac{1}{n-1}\sum_{j=1,j \ne i}^n \int_{\delta}^t  \mathbb{E}\bigg[\Big|(\br_s^j-\bar{A}^n_{s})\left(f^P(\r_{s-\delta}^{j,n}-A_{s-\delta}^n)-f^P(\br_{s-\delta}^{j}-\bar{A}_{s-\delta}^n)\right)\Big|\bigg]ds \notag \\ 
& \quad \le K_2 \int_{\delta}^t  \mathbb{E}[|\r^{i,n}_s-\br_s^i|]ds +K_2 \int_{\delta}^t  \mathbb{E}[|A^{n}_s-\bar{A}^n_s|ds  \notag \\
&\qquad + \int_{\delta}^t  \mathbb{E}\left[\left|\br_s^i-\bar{A}^n_{s}\right|\left|f^P(\r_{s-\delta}^{i,n}-A_{s-\delta}^n)-f^P(\br_{s-\delta}^{i}-\bar{A}_{s-\delta}^n)\right|\right]ds. 
\end{align}
We have that for $\delta \le t$
\begin{align}
&\int_{\delta}^t  \mathbb{E}\left[\left|\br_s^i-\bar{A}^n_{s}\right|\left|f^P(\r_{s-\delta}^{i,n}-A_{s-\delta}^n)-f^P(\br_{s-\delta}^{i}-\bar{A}_{s-\delta}^n)\right|\right]ds \notag \\ 
& \quad \le \int_{\delta}^t  \left(\mathbb{E}\big[\big|\br_s^i-\bar{A}_s^n\big|^2\big]ds\right)^{1/2}  \left(\mathbb{E}\left[\left|f^P(\r_{s-\delta}^{i,n}-A^n_s)-f^P(\br_{s-\delta}^{i}-\bar{A}_s^n)\right|^2\right]\right)^{1/2}ds\notag \\
&\quad \le\left(\int_{\delta}^t \mathbb{E}\left[\left|\br_s^i-\bar{A}^n_{s}\right|^2\right] ds\right)^{1/2}\left(\int_{\delta}^t \mathbb{E}\left[\left|f^P(\r_{s-\delta}^{i,n}-A^n_{s-\delta})-f^P(\br_{s-\delta}^{i}-\bar{A}_{s-\delta}^n)\right|^2\right]ds\right)^{1/2}\notag 
\end{align}
\begin{align}
&  \quad \le \left(\int_{\delta}^t \mathbb{E}\left[\left|\br_s^i-\bar{A}^n_{s}\right|^2\right]ds \right)^{1/2} \left( \int_{\delta}^t  \mathbb{E}\left[\left|f^P(\r_{s-\delta}^{i,n}-A^n_{s-\delta})^2-f^P(\br_{s-\delta}^{i}-\bar{A}^n_{s-\delta})^2\right|\right]ds\right)^{1/2} \notag \\
 & \quad  \le  \sqrt{2K_2}\left(\int_{\delta}^t \mathbb{E}\left[\left|\br_s^i-\bar{A}^n_{s}\right|^2\right]ds\right)^{1/2}\left(\int_{\delta}^t \mathbb{E}\left[\left|f^P(\r_{s-\delta}^{i,n}-A^n_{s-\delta})-f^P(\br_{s-\delta}^{i}-\bar{A}_{s-\delta}^n)\right|\right]ds\right)^{1/2} \notag \\
  & \quad  \le \sqrt{2K_1K_2}\left(\int_{\delta}^t \mathbb{E}\left[\left|\br_s^i-\bar{A}^n_{s}\right|^2\right]ds \right)^{1/2}\left(\int_{\delta}^t \mathbb{E}\left[\left|\r_{s-\delta}^{i,n}-\br_{s-\delta}^{i}\right|+\left|A_{s-\delta}^n-\bar{A}_{s-\delta}^n \right|\right]ds\right)^{1/2},  \quad  \notag
\end{align}
where we have used that $|a-b|^2 \le |a^2 - b^2|$ for $a,b \in \mathbb{R}^+$.\\
Then, setting $G_1^n(t):=\left(\int_{\delta}^t \mathbb{E}\left[\left|\br_s^i-\bar{A}^n_{s}\right|^2\right] ds\right)^{1/2},$ by (\ref{int11}) we have 
\begin{align}\label{int111}
 &\int_{\delta}^t \mathbb{E}\bigg[\Big|\frac{1}{n-1} \sum_{j=1,j \ne i}^n (f^P(\r_{s-\delta}^{j,n}-A^n_{s-\delta})(\r^{j,n}_s-A^n_{s}) - f^P(\br^j_{s-\delta}-\bar{A}^n_{s-\delta})(\br_s^j-\bar{A}_s^n))\Big|\bigg]ds \notag \\
&  \le  K_2 \int_{\delta}^t  \mathbb{E}[|\r^{i,n}_s-\br_s^i|]ds +K_2 \int_{\delta}^t  \mathbb{E}[|A^{n}_s-\bar{A}^n_s|ds \notag \\ & \quad+ \sqrt{2K_1K_2}G_1^n(t) \left(\int_{\delta}^t \mathbb{E}\left[\left|\r_{s-\delta}^{i,n}-\br_{s-\delta}^{i}\right|+\left|A_{s-\delta}^n-\bar{A}_{s-\delta}^n \right|\right]ds\right)^{1/2}, \quad  \delta \le t.
\end{align}
Since 
\begin{align}
&\int_{\delta}^t \mathbb{E}\left[\left|\r_{s-\delta}^{i,n}-\br_{s-\delta}^{i}\right|+\left|A_{s-\delta}^{n}-\bar{A}_{s-\delta}^{n}\right|\right]ds 
= \mathbb{E}\left[\int_{\delta}^t \left(\left|\r_{s-\delta}^{i,n}-\br_{s-\delta}^{i}\right|+\left|A_{s-\delta}^{n}-\bar{A}_{s-\delta}^{n}\right|\right) ds\right]\notag \\ &\quad= \mathbb{E}\left[\int_{0}^{t-\delta}\left(\left|\r_{u}^{i,n}-\br_{u}^{i}\right|+\left|A_{u}^{n}-\bar{A}_{u}^{n}\right|\right)du\right]  \le \int_0^{\delta}\mathbb{E}[ \left|\r_{u}^{i,n}-\br_{u}^{i}\right|+\left|A_{u}^{n}-\bar{A}_{u}^{n}\right|]du, \quad \delta \le t < 2\delta, \notag
\end{align}
we can rewrite (\ref{int111}) as
\begin{align}\label{int1}
 &\int_{\delta}^t \mathbb{E}\bigg[\Big|\frac{1}{n-1} \sum_{j=1,j \ne i}^n (f^P(\r_{s-\delta}^{j,n}-A^n_{s-\delta})(\r^{j,n}_s-A^n_{s}) - f^P(\br^j_{s-\delta}-\bar{A}^n_{s-\delta})(\br_s^j-\bar{A}^n_s))\Big|\bigg]ds \notag \\
 &  \le  K_2 \int_{\delta}^t  \mathbb{E}[|\r^{i,n}_s-\br_s^i|]ds +K_2 \int_{\delta}^t  \mathbb{E}[|A^{n}_s-\bar{A}^n_s|ds \notag \\ 
 & \quad+ \sqrt{2K_1K_2}G_1^n(t) \left(\int_0^{\delta} \mathbb{E}\left[\left|\r_{s}^{i,n}-\br_{s}^{i}\right|+\left|A_{s}^n-\bar{A}_{s}^n \right|\right]ds\right)^{1/2}, \quad  \delta \le t.
\end{align}

Similarly,
\begin{align}\label{int2}
&  \int_0^t\mathbb{E}\bigg[\Big| \frac{1}{m}\sum_{k=1}^m  \left(f^B(\r_{s-\delta}^{k,B}-A_{s-\delta}^n)(\r^{k,B}_s-A_s^n)-f^B(\br_{s-\delta}^{k,B}-\bar{A}_{s-\delta}^n)(\br^{k,B}_s-\bar{A}_s^n)\right)\Big|\bigg]ds \notag \\
& \quad \le  K_2 \int_{\delta}^t  \mathbb{E}[|\r^{k,B}_s-\br_s^{k,B}|]ds +K_2 \int_{\delta}^t  \mathbb{E}[|A^{n}_s-\bar{A}^n_s|ds \notag \\ 
& \quad \quad+ \sqrt{2K_1K_2}G_2^n(t) \left(\int_0^{\delta} \mathbb{E}\left[\left|\r_{s}^{k,B}-\br_{s}^{k,B}\right|+\left|A_{s}^n-\bar{A}_{s}^n \right|\right]ds\right)^{1/2}, \quad  \delta \le t.
\end{align}
with $G_2^n(t):=\left(\int_{\delta}^t \mathbb{E}\left[\left|\br_s^{k,B}-\bar{A}^n_{s}\right|^2\right]ds \right)^{1/2}.$\\
From (\ref{A}), (\ref{ledelte}), (\ref{diff}), (\ref{int1}) and (\ref{int2}) we obtain
\begin{align}
 & \mathbb{E}[|\rho^{i,n}-\br^i|_t^*] \notag \\
 &\le  (3K_2+2) \int_{\delta}^t  \mathbb{E}\left[\left|\r^{i,n} - \br^i\right|_s^*\right]ds + (3K_2+1) \int_{\delta}^t  \mathbb{E}[|\r^{k,B} - \br^{k,B}|_s^*]ds  \notag \\
& \quad + \sqrt{2K_1K_2}G_1^n(t) \left(\int_0^{\delta} \mathbb{E}\left[\left|\r_{s}^{i,n}-\br_{s}^{i}\right|+\left|A_{s}^n-\bar{A}_{s}^n \right|\right]ds\right)^{1/2} \notag \\
& \quad+ \sqrt{2K_1K_2}G_2^n(t) \left(\int_0^{\delta} \mathbb{E}\left[\left|\r_{s}^{k,B}-\br_{s}^{k,B}\right|+\left|A_{s}^n-\bar{A}_{s}^n \right|\right]ds\right)^{1/2} \notag \\
  &  \quad + \int_{0}^t \mathbb{E}\bigg[\left|f^B(\bar{\r}_{s-\delta}^{k,B}-\bar{A}_{s-\delta}^n)(\bar{\r}^{k,B}_s-\bar{A}_s^n)-f^B(\br_{s-\delta}^{k,B}-\nu_{s-\delta}-\mathbb{E}[\tilde{\rho}^i_{s-\delta}])(\br^{k,B}_s-\nu_s-\mathbb{E}[\tilde{\rho}^i_s])\right|\bigg]ds \notag \\
 &\quad + \int_{0}^t \mathbb{E}\bigg[\Big|\frac{1}{n-1}\sum_{j=1,j\ne i}^n f^P(\bar{\rho}_{s-\delta}^{j}-\bar{A}_{s-\delta}^n) (\bar{\rho}_s^{j}-\bar{A}_s^n)-  \mathbb{E}\left[f^P\left(\tr_{s-\delta}^i-\mathbb{E}[\tr_{s-\delta}^i]\right)\left(\tr^i_{s}-\mathbb{E}[\tr^i_s]\right)\right]\Big|\bigg]ds \notag \\
 &\quad+ \int_0^{\delta} \mathbb{E}[|\r^{i,n}_s-\br_s^i|]ds+\int_0^t\mathbb{E}\left[|\bar{A}_s^n-\mathbb{E}[\tilde{\rho}^i_s]-\nu_s|\right]ds, \quad  \delta \le t <2\delta \label{eqi}. 
\end{align}
At the same way, by (\ref{rhobubble}) and (\ref{barrhobubble}) we have
\begin{align}
 & \mathbb{E}[|\rho^{k,B}-\br^{k,B}|_t^*] \notag \\
 &\le  (3K_2+1) \int_{\delta}^t  \mathbb{E}\left[\left|\r^{i,n} - \br^i\right|_s^*\right]ds + (3K_2+2) \int_{\delta}^t  \mathbb{E}[|\r^{k,B} - \br^{k,B}|_s^*]ds  \notag \\
& \quad + \sqrt{2K_1K_2}G_1^n(t) \left(\int_0^{\delta} \mathbb{E}\left[\left|\r_{s}^{i,n}-\br_{s}^{i}\right|+\left|A_{s}^n-\bar{A}_{s}^n \right|\right]ds\right)^{1/2} \notag \\
& \quad+ \sqrt{2K_1K_2}G_2^n(t) \left(\int_0^{\delta} \mathbb{E}\left[\left|\r_{s}^{k,B}-\br_{s}^{k,B}\right|+\left|A_{s}^n-\bar{A}_{s}^n \right|\right]ds\right)^{1/2} \notag \\
  &  \quad + \int_{0}^t \mathbb{E}\bigg[\left|f^B(\bar{\r}_{s-\delta}^{k,B}-\bar{A}_{s-\delta}^n)(\bar{\r}^{k,B}_s-\bar{A}_s^n)-f^B(\br_{s-\delta}^{k,B}-\nu_{s-\delta}-\mathbb{E}[\tilde{\rho}^i_{s-\delta}])(\br^{k,B}_s-\nu_s-\mathbb{E}[\tilde{\rho}^i_s])\right|\bigg]ds \notag \\
 &  \quad + \int_{0}^t \mathbb{E}\bigg[\Big|\frac{1}{n}\sum_{i=1}^n f^P(\bar{\rho}_{s-\delta}^{i}-\bar{A}_{s-\delta}^n) (\bar{\rho}_s^{i}-\bar{A}_s^n)-  \mathbb{E}\left[f^P\left(\tr_{s-\delta}^i-\mathbb{E}[\tr_{s-\delta}^i]\right)\left(\tr^i_{s}-\mathbb{E}[\tr^i_s]\right)\right]\Big|\bigg]ds \notag \\
 & \quad + \int_0^{\delta} \mathbb{E}[|\r^{k,B}_s-\br_s^{k,B}|]ds+ \int_0^t \mathbb{E}\left[|\bar{A}_s^{n}-\nu_s-\mathbb{E}[\tilde{\nu}^i_s|]\right]ds, \quad  \delta \le t <2\delta \label{eqk}. 
\end{align}

Summing up (\ref{eqi}) and (\ref{eqk}) we find 
\begin{align}
&\mathbb{E}[|\r^{i,n}-\br^i|^*_t]+\mathbb{E}[|\r^{k,B}-\br^{k,B}|_t^*] \notag \\
& \quad \le (6K_2+3)\int_{0}^t  \left(\mathbb{E}[|\r^{i,n}-\br^i|_s^*]+\mathbb{E}[|\r^{k,B} - \br^{k,B}|_s^*]\right)ds\notag \\ 
&  \quad +\sqrt{2K_1K_2}(G^n_1(t)+G^n_2(t))\left(\int_{0}^{\delta} \left(\mathbb{E}[|\r_s^{i,n}-\br_s^i|]+\mathbb{E}[|\r_s^{k,B} - \br_s^{k,B}|]+\mathbb{E}\left[\left|A_s^n-\bar{A}^n_s\right|\right]\right)ds
\right)^{1/2}\label{finale2} \\
& \quad+ \int_{0}^t \mathbb{E}\bigg[\left|f^B(\bar{\r}_{s-\delta}^{k,B}-\bar{A}_{s-\delta}^n)(\bar{\r}^{k,B}_s-\bar{A}_s^n)-f^B(\br_{s-\delta}^{k,B}-\nu_{s-\delta}-\mathbb{E}[\tilde{\rho}^i_{s-\delta}])(\br^{k,B}_s-\nu_s-\mathbb{E}[\tilde{\rho}^i_s])\right|\bigg]ds \notag \\
 &\quad + \int_{0}^t \mathbb{E}\bigg[\Big|\frac{1}{n-1}\sum_{j=1,j\ne i}^n f^P(\bar{\rho}_{s-\delta}^{j}-\bar{A}_{s-\delta}^n) (\bar{\rho}_s^{j}-\bar{A}_s^n)-  \mathbb{E}\left[f^P\left(\tr_{s-\delta}^i-\mathbb{E}[\tr_{s-\delta}^i]\right)\left(\tr^i_{s}-\mathbb{E}[\tr^i_s]\right)\right]\Big|\bigg]ds \notag \\
 &  \quad  + \int_{0}^t \mathbb{E}\bigg[\Big|\frac{1}{n}\sum_{i=1}^n f^P(\bar{\rho}_{s-\delta}^{i}-\bar{A}_{s-\delta}^n) (\bar{\rho}_s^{i}-\bar{A}_s^n)-  \mathbb{E}\left[f^P\left(\tr_{s-\delta}^i-\mathbb{E}[\tr_{s-\delta}^i]\right)\left(\tr^i_{s}-\mathbb{E}[\tr^i_s]\right)\right]\Big|\bigg]ds \notag \\
 & \quad+ 2\int_0^t \mathbb{E}\left[|\bar{A}_s^{n}-\nu_s-\mathbb{E}[\tilde{\nu}^i_s|]\right]ds, \quad \delta \le t <2\delta. \label{finale}
\end{align}
With the same computations used in the first step of the proof, we show that the last four terms of (\ref{finale}) converge to zero when $n \to \infty$ by the proof of Proposition \ref{lemmalim}. The term in (\ref{finale2}) also goes to zero when $n \to \infty$, by the first step of the proof and because $\lim_{n \to \infty} [G^n_1(t)+G^n_2(t)]<\infty$, by (\ref{footnote}). Then applying Gronwall's Lemma to (\ref{finale}) we prove the result for $t \in [\delta,2\delta)$. \\
The result then follows by proceeding in the same way for all the steps $t \in [k \delta,(k+1)\delta),$ $k \ge 2$. $\quad \Box$

\section{Numerical analysis}\label{simulations}
We now study by numerical simulations how the system described in Section \ref{seclimit} reacts to the growth and the burst of a bubble. 
In particular, we investigate how a bank not holding the bubbly asset can be affected by a bubble burst through contagion mechanisms. We first consider the case of \eqref{rhonobubble}-\eqref{rhobubble}, i.e. of a network with a finite number of banks, and then we analyze the limit system (\ref{rhotilde})-(\ref{barrhobubble}). 
\\
The bubble has the dynamics specified in \citet{firstpaper}, i.e. it solves (\ref{dbeta}) with
$$
\mu_t=M_t\Lambda_t(-k\beta_t+2\bar{\mu}_t), \qquad \sigma_t=2\bar{\sigma}M_t\Lambda_t, \quad t \ge 0,
$$
 where $M=(M_t)_{t \in [0, T]}$, $\Lambda=(\Lambda_t)_{t \in [0, T]}$ are respectively a measure of illiquidity and the so called \textit{resiliency},  $\bar{\mu}=(\bar{\mu}_t)_{t \ge 0}$ is the drift of the signed volume of market orders (buy market orders minus sell market orders) and $\bar{\sigma}>0$. 
Here, the illiquidity $M$ is supposed to be a geometric Brownian motion, i.e.
$$
dM_t=M_t(\mu^Mdt+\sigma^MdB_t^3), \quad t \ge 0.
$$
with $\mu^M \in \mathbb{R}$ and $\sigma^M>0$. We choose the same function $f$ for both core and periphery banks in (\ref{rhonobubble})-(\ref{rhobubble}), i.e. $f^B=f^P=f$. In particular, we take $f(x)=1+2\arctan(x)/\pi$, as in Example \ref{exf}. 

\subsection{Risk analysis for the finite case}

We first focus on the system (\ref{rhonobubble})-(\ref{rhobubble}). We investigate how the first bank reacts when banks holding the bubble are in trouble. Specifically, we here introduce and compute the risk measure
\begin{equation}\label{riski}
Risk^i_{\alpha}=-\sup_{x \in \mathbb{R}} \left\{ \left[\frac{1}{N_s}\sum_{k=1}^{N_s}\mathbb{1}_{\left\{(\rho_{\tau_k+\Delta}^{i,n,k}-\rho^{i,n,k}_{\tau_k})/\rho^{i,n,k}_{\tau_k} \le x\right\}}\right] \le \alpha \right\},
\end{equation}
with $\alpha>0$, where $N_s$ is the number of simulations of the processes in (\ref{rhonobubble})-(\ref{rhobubble}), $\tau_k$ is the value at the $k$-th simulation of the bursting time $\tau$ of the bubble,  and $\rho^{i,n,k}_{t}$ is the value of $\rho_t^{i,n}$ computed in the $k$-th simulation.  \\ The risk measure $Risk_{\alpha}^i$ as defined in (\ref{riski}) is  analogous to the CoVar of a bank without the bubble with respect to a bank with the bubble (for a definition of CoVar see e.g. \citet{biaginisyst} and \citet{Brunnermeier2012}). Note that, since the banks not holding the bubble are identically distributed, we only compute the risk for one bank. \\
From now on, we set $\alpha=0.05$ in \eqref{riski}. We perform $N_s=10000$ simulations of $Risk^1_{0.05}$ in the case when there are $n=6$ banks not holding the bubble and $m=2$ banks holding it. We consider different values of $\lambda$ and of the delay $\delta$. \\ The results are given in Table \ref{rhominumeanbubble}: \\
\begin{table}[h]
 \centering
 \begin{tabular}{|c|c|c|c|c|c|c|c|c|}
 \hline
& $\delta=0$ & $\delta=0.025$ & $\delta=0.05$ &  $\delta=0.075$ &  $\delta=0.1$ &  $\delta=0.2$ & $\delta=0.3$\\
 \hline
  $\lambda=0.5$  & $0.283$ & $0.390$ & $0.451$ &  $0.716$  &   $0.925$& $0.916$  & $0.901$   \\
 \hline
 $\lambda=1$ & $0.281$ & $0.385$ & $0.434$ &  $0.661$  &   $0.886$& $0.879$ &  $0.875$ \\
\hline
$\lambda=2$ & $0.280$ & $0.377$ & $0.422$ &  $0.641$  &   $0.851$ & $0.824$   & $0.819$   \\
 \hline
\end{tabular}
 \caption{$Risk^1_{0.05}$ in the case when the robustness is given by (\ref{rhonobubble})-(\ref{rhobubble}), with parameters $\sigma_1=\sigma_2=0.2$, $\Delta=0.1$, $\rho_0^{i,6}=\rho^{k,B}_0=0.5$, $i=1,\dots,6$, $k=1,2$. }
 \label{rhominumeanbubble}
\end{table}\\
As expected, the risk is bigger for large delays, since a large delay means that the banks without the bubble are not able to quickly disinvest,  when  other institutions holding the bubble are in trouble. However, for delays larger than $0.1$, the risk is still big but it decreases. This depends on the fact that we check the robustness of the banks at time $\tau+0.1$: at this time, when $\delta=0.2,0.3$, $f$ is smaller than in the case $\delta=0.1$ because banks are cross investing on each other according to a value of the robustness, which is realized much before the bubble's burst. \\ Moreover, the risk is decreasing with $\lambda$. Indeed, it follows by (\ref{rhonobubble}) that $\rho^{i,n}$ reverts to 
$$ 
A_t^n+\frac{1}{\lambda}\left(\frac{1}{n}\sum_{i=1}^n f(\rho_{t-\delta}^{i,n}-A^n_{t-\delta}) (\rho_t^{i,n}-A_t^n) + \frac{1}{m-1} \sum_{\ell=1,\ell \ne k}^m  f(\rho_{t-\delta}^{\ell,B}-A^n_{t-\delta}) (\rho_t^{\ell,n}-A_t^n)\right),
$$
so that for large $\lambda$ the term involving the network, and then the direct effects of the banks holding the bubbly asset, is less significative.

We now consider (\ref{rhonobubble})-(\ref{rhobubble}) when $\beta$ is replaced by $\bar{\beta}$, where
\begin{align}\label{barbeta}
d\bar{\beta}_t=
\begin{cases}
0 \quad  \text{for } t \le \tau, \\
\frac{\rho^{1,\bar{\beta}}_{\tau}}{\rho^{1,\beta}_{\tau}}d\beta_t   \quad  \text{for } t \ge \tau, 
\end{cases}
\end{align}
where $\rho^{1,\beta}$ is the robustness of bank 1 when there is a bubble in the network, and $\rho^{1,\bar{\beta}}$ is the robustness of bank 1 when there is no bubble. In this way we model the case when the banks that used to hold the bubbly asset are subject at time $\tau$ to the same (relative) shock, but without having experienced the growth of the bubble. The results are given in Table \ref{rhominumeannobubble}, for the same parameters as in Table \ref{rhominumeanbubble}.  
\begin{table}[h]
 \centering
 \begin{tabular}{|c|c|c|c|c|c|c|c|}
 \hline
& $\delta=0$ & $\delta=0.025$ & $\delta=0.05$ &  $\delta=0.075$ &  $\delta=0.1$ &  $\delta=0.2$  & $\delta=0.3$\\
 \hline
 $\lambda=0.5$  &  $0.281$ & $0.383$ & $0.388$ &    $0.415$ & $0.505$ & $0.499$  & $0.494$  \\
 \hline
$\lambda=1$ & $0.280$ & $0.381$ & $0.385$ &  $0.403$  &   $0.502$ & $0.494$ & $0.492$ \\
 \hline
$\lambda=2$ & $0.280$ & $0.371$ & $0.380$ &  $0.399$ &  $0.500$ & $0.490$ & $0.489$     \\
\hline
\end{tabular}
 \caption{$Risk^1_{0.05}$ in the case when the robustness is given by (\ref{rhonobubble})-(\ref{rhobubble}) with no bubble in the system, but with the same shock at time $\tau$, for parameters $\sigma_1=\sigma_2=0.2$, $\Delta=0.1$, $\rho_0^{i,6}=\rho^{k,B}_0=0.5$, $i=1,\dots,6$, $k=1,2$.}
 \label{rhominumeannobubble}
\end{table}\\

We note that for $\delta=0$ there is not any significant difference with the case when there is a bubble in the system, since the banks are able to disinvest immediately at the time when the shock hits the banks with the bubble. Anyway, this difference increases with the delay. When the delay is big, the banks with no bubble are much more in trouble in the first case, i.e when they are attached to banks holding the bubbly asset. \\  We can then conclude that the increase of the value of the bubbly asset can put the network in trouble, because it makes the system more  centralized on the riskier banks, due to the preferential attachment mechanism implied by (\ref{rhonobubble})-(\ref{rhobubble}).\\
This can also be seen by considering a static network, i.e. by taking $f^B=f^P = 1$ in (\ref{rhonobubble})-(\ref{rhobubble}). In this case, we obtain the following values of the risk for different values of $\lambda$:
 \begin{table}[h]
 \centering
 \begin{tabular}{|c|c|c|}
 \hline
  $\lambda=0.5$ & $\lambda=1$ & $\lambda=2$ \\
 \hline
  $0.670$ & $0.626$ & $0.599$ 		\\
\hline
\end{tabular}
 \caption{$Risk^1_{0.05}$ with $\Delta=0.1$ in the case of a static network, with $f^B = f^P = 1$ and with parameters $\sigma_1=\sigma_2=0.2$, $\Delta=0.1$, $\rho_0^{i,6}=\rho^{k,B}_0=0.5$, $i=1,\dots,6$, $k=1,2$..}
 \label{rhominumeanstatic}
\end{table}\\
Note that in this case the delay plays no role since it only affects the dynamics through $f^B$ and $f^P$. 
Comparing this result with Table \ref{rhominumeanbubble}, one can see that when $\delta$ in (\ref{rhonobubble})-(\ref{rhobubble}) is small, then the fact that banks are able to quickly disinvest  makes the system safer than in the case of a static network. On the other hand, for big values of $\delta$, a centralized network towards the banks holding the bubble and the impossibility to disinvest quickly after the burst give rise to a more dangerous system than in the static case. \\
\subsection{Risk analysis for the mean field limit}
We now consider the case of the limit system (\ref{rhotilde})-(\ref{barrhobubble}). We compute
\begin{equation}\label{riski1}
Risk^1_{0.05}=-\sup_{x \in \mathbb{R}} \left\{ \left[\frac{1}{N_s}\sum_{k=1}^{N_s}\mathbb{1}_{\left\{(\bar{\rho}_{\tau_k+\Delta}^{1,k}-\bar{\rho}^{1,k}_{\tau_k})/\bar{\rho}^{1,k}_{\tau_k} \le x\right\}}\right] \le 0.05 \right\},
\end{equation}
where $N_s$ and $\tau_k$ are the number of simulations and the time of the burst of the bubble in the $k$-th simulation, respectively, and $\bar{\rho}_t^{1,k}$ is the value of $\bar{\rho}^1_t$ computed in the $k$-th simulation. \\
As before, we consider $m=2$ banks holding the bubble  
 and we make $N_s=10000$ simulations of (\ref{rhotilde})-(\ref{barrhobubble}) taking different values of $\lambda$ and  $\delta$. \\ 
 We compute $\phi(t,t-\delta)=\mathbb{E}\left[f\left(\tr_{t-\delta}^i-\mathbb{E}[\tr_{t-\delta}^i]\right)\left(\tr_{t}^i-\mathbb{E}[\tr_t^i]\right)\right]$ in (\ref{nu}) and (\ref{barrhobubble}) via Monte Carlo simulations of the trajectories of the Ornstein-Uhlenbeck process in (\ref{rhotilde}). Note that $\mathbb{E}[\tilde{\rho}^i_t]=\rho_0e^{-\lambda t}.$ 
The results are gathered in Table \ref{rhominumeannobubblelimit}.
\begin{table}[h]
 \centering
 \begin{tabular}{|c|c|c|c|c|c|c|c|c|}
 \hline
& $\delta=0$ & $\delta=0.025$ & $\delta=0.05$ &  $\delta=0.075$ &  $\delta=0.1$ &  $\delta=0.2$ & $\delta=0.3$\\
  \hline
   $\lambda=0.5$  & $0.305$ & $0.367$ & $0.563$ &  $0.908$  &   $1.281$& $1.251$  & $1.226$   \\
 \hline
 $\lambda=1$ & $0.302$ & $0.360$ & $0.521$ &  $0.765$  &   $1.170$ & $1.125$  & $1.117$ \\
\hline
$\lambda=2$ & $0.302$ & $0.356$ & $0.503$ &  $0.647$  &   $0.908$ & $0.907$   & $0.877$   \\
 \hline
\end{tabular}
 \caption{$Risk^1_{0.05}$ with $\Delta=0.1$ of the mean field limit (\ref{rhotilde})-(\ref{barrhobubble}), with parameters $\sigma_1=\sigma_2=0.2$,  $\rho^{k,B}_0=0.5$, $k=1,2$. }
 \label{rhominumeannobubblelimit}
\end{table}\\
As before, the risk is increasing with the delay until $\delta=0.1$ and decreasing with $\lambda$, since $\bar{\rho}_t^i$ reverts to 
$$
\frac{1}{\lambda}\left( \varphi(t,t-\delta)+\frac{1}{m} \sum_{k=1}^m f\left(\br_{t-\delta}^{k,B}-\nu_{t-\delta}-\mathbb{E}[\tr^i_{t-\delta}]\right)\left(\br_t^{k,B}-\nu_t-\mathbb{E}[\tr^i_t]\right)\right)+\mathbb{E}[\tilde{\rho}^i_t]-\tilde{\rho}^i_t,
$$
so that a large $\lambda$ diminishes the influence of the banks holding the bubbly asset. \\
We can also see that the risk is bigger at the limit by comparing \eqref{rhonobubble} and \eqref{barrhonobubble}: since $\nu_{t-\delta}+\mathbb{E}[\tilde{\rho}^i_t]<A_{t-\delta}^n$,  because the first term is the average robustness of banks not holding the bubble, the argument of $f$ is bigger in \eqref{barrhonobubble}. This leads to a bigger weight multiplying the loss at the moment of the burst at the limit. \\
In  Table \ref{rhominusmeannobubblelimit}, we report the results for the case when $\beta$ is replaced by $\bar{\beta}$ as in (\ref{barbeta}), i.e. when there is no bubble in the network. 
 \begin{table}[h]
 \centering
 \begin{tabular}{|c|c|c|c|c|c|c|c|}
 \hline
& $\delta=0$ & $\delta=0.025$ & $\delta=0.05$ &  $\delta=0.075$ &  $\delta=0.1$ &  $\delta=0.2$ & $\delta=0.3$ \\
 \hline
 $\lambda=0.5$  &$0.303$ & $0.355$ & $0.468$ &  $0.682$  &   $0.698$ & $0.659$ & $0.645$   \\
  \hline
$\lambda=1$ & $0.302$ & $0.347$ & $0.410$ &  $0.528$  &  $0.640$ & $0.628$  & $0.627$  \\
 \hline
$\lambda=2$ & $0.300$ & $0.340$ & $0.395$ &  $0.455$ &  $0.612$ & $0.561$ & $0.550$     \\
\hline
\end{tabular}
 \caption{$Risk^1_{0.05}$ with $\Delta=0.1$ in the mean field limit  (\ref{rhotilde})-(\ref{barrhobubble}) with no bubble, with parameters $\sigma_1=\sigma_2=0.2$, $\rho^{k,B}_0=0.5$, $k=1,2$.}
 \label{rhominusmeannobubblelimit}
\end{table}\\
As before, it can be seen that, when the delay is large enough, the preferential attachment mechanism, that takes place during the ascending phase of the bubble, creates a network more exposed to systemic risk at the time of the shock. If we consider a a static network, with $f^B=f^P=1$, the results, shown in Table \ref{rhominumeanstaticlimit}, agree with the ones obtained in the case of the finite network: for small delays the dynamic network is less exposed to systemic risk with respect to the static one, whereas when the delay increases and the banks in the dynamic network are slower in disinvesting, the risk is bigger than for the static network.
 \begin{table}[h]
 \centering
 \begin{tabular}{|c|c|c|}
 \hline
  $\lambda=0.5$ & $\lambda=1$ & $\lambda=2$ \\
 \hline
  $1.001$ & $0.910$ & $0.866$ 		\\
\hline
\end{tabular}
 \caption{$Risk^1_{0.05}$ with $\Delta=0.1$ in the case of a static network with $f^B = f^P = 1$ in the mean field limit, with parameters $\sigma_1=\sigma_2=0.2$, $\Delta=0.1$,  $\rho^{k,B}_0=0.5$, $k=1,2$.}
 \label{rhominumeanstaticlimit}
\end{table}

 \appendix
 
 \section{Existence and uniqueness theorems}
 
 For the reader's convenience we report here the results, which we have used in the paper to prove existence and uniqueness of a strong solution of a system of stochastic differential equations (SDEs) and of stochastic differential delay equations (SDDEs). These theorems also guarantee the finiteness of the second moments of the strong solution. \\
 In the following, let $(\Omega,\cal{F},P)$ be a complete probability space with a filtration $\mathbb{F}:=(\cal{F}_t)_{t \ge 0}$ satisfying the usual conditions, and $B_t = (B^1_t, \dots , B^m_t)_{t \ge 0}$, be an $m$-dimensional $\mathbb{F}$-Brownian motion defined on $(\Omega,\cal{F},P)$.  
 \\ We begin by the following existence and uniqueness result for a system of SDEs, given in Theorem 9.11 in \citet{Pascucci2011}. 
\begin{theorem}
Let $X_0$ be an $\cal{F}_{t_0}$-measurable $\mathbb{R}^d$-valued random variable such that $\mathbb{E}[X_0^2] <\infty$. Consider the $d$-dimensional stochastic differential equation of It\^o  type
\begin{equation}\label{eqtheo1}
dX_t=f(t,X_t)dt+g(t,X_t)dB_t, \quad t_0 \le t \le T,  
\end{equation}
with $X_{t_0}=X_0$, where $f:[t_0,T]\times \mathbb{R}^d\to \mathbb{R}^d$ and $g:[t_0,T] \times \mathbb{R}^d \to \mathbb{R}^{d\times m}$ are both Borel measurable. \\
Assume that there there exist two positive constants $K_1$ and $K_2$ such that:
\begin{enumerate}
\item (Lipschitz condition) for all $x, y \in \mathbb{R}^d$ and $t \in [t_0, T ]$,  
$$
\lVert f(t,x)-f(t,y)\rVert^2+\lVert g(t,x)-g(t,y)\rVert^2 \le K_1\lVert x-y\rVert^2;
$$
\item (Linear growth condition) for all $(t,x) \in [t_0, T] \times \mathbb{R}^d  $, 
$$
\lVert f(t,x)\rVert^2+\lVert g(t,x)\rVert^2 \le K_2(1+\lVert x\rVert^2). 
$$
\end{enumerate}
Then there exists a unique solution $X=(X_t)_{x \in [t_0,T]}$ to equation $(\ref{eqtheo1})$ and it holds
\begin{equation}\label{pascucci}
\mathbb{E}\left[\sup_{t_0 \le s \le t} \lVert X_s\rVert^2 \right] \le C(1+\mathbb{E}\left[\lVert X_0 \rVert^2 \right])e^{Ct}, \quad t \in [t_0,T],
\end{equation}
where $C$ is a constant depending on $K_2$ and $T$ only.
\end{theorem}
 
 We now recall Theorem 3.1 in \citet[chapter 5]{Mao}, that provides the existence and uniqueness results for SDDEs. 
 
 \begin{theorem}
 Let $F:  [t_0,T]\times \mathbb{R}^d \times \mathbb{R}^d \to \mathbb{R}^d$ and
$G : [t_0,T] \times\mathbb{R}^d\times \mathbb{R}^d \to \mathbb{R}^{d\times m}$ be Borel-measurable. Consider the delay equation
 \begin{equation}\label{eqtheo2}
 dX_t = F(t,X_t, X_{t - \tau})dt + G(t,X_t, X_{t - \tau})dB_t, 
 \end{equation}
 with initial data $\{X_s: t_0-\tau\le s \le t_0\}$, such that $X_s$ is $\cal{F}_{t_0}$-measurable for all $s \in [t_0-\tau,t_0]$ and $\mathbb{E}[\lVert X_s \lVert^2] < \infty$ for all $s \in [t_0-\tau,t_0]$. \\
 Assume that there exists two positive constants $\tilde{K_1}$ and $\tilde{K_2}$ such that
 \begin{enumerate}
 \item (Linear growth condition) for all $(t,x,y) \in [t_0, T]  \times \mathbb{R}^d \times \mathbb{R}^d  $, 
 $$
 \lVert F(t,x,y)\rVert ^2+\lVert G(t,x, y)\rVert^2\le \tilde{K_1}(1+\lVert x\rVert^2+\lVert y \rVert^2);
 $$
 \item (Lipschitz condition on $x$) for all $t\in [t_0,T]$, $y\in \mathbb{R}^d$ and $x,\bar{x}\in \mathbb{R}^d$, 
 $$
 \lVert F(t,x,y)-F(t,\bar{x},y)\rVert^2+\lVert G(t,x,y)-G(t,\bar{x},y)\rVert^2 \le \tilde{K_2}\lVert x-\bar{x} \rVert^2.
$$
\end{enumerate}
Then there exists a unique solution $X=(X_t)_{t \in [t_0,T]}$ to equation $(\ref{eqtheo2})$. 
  \end{theorem}

\bibliography{bib}
\bibliographystyle{plainnat}
\end{document}